\documentclass{article}
\usepackage{float}
\usepackage{gensymb}
\def\arcmin{\ensuremath{{}^{\prime}}}

\usepackage{threeparttable}

\usepackage{graphicx}
\usepackage{caption}
\usepackage{subcaption} 
\usepackage{savesym}
\savesymbol{iint}
\savesymbol{iiint}
\savesymbol{iiiint}
\savesymbol{idotsint}
\savesymbol{leftroot}
\savesymbol{uproot}
\usepackage{amsmath}
\restoresymbol{AMS}{iint}
\restoresymbol{AMS}{iiint}
\restoresymbol{AMS}{iiiint}
\restoresymbol{AMS}{idotsint}
\restoresymbol{AMS}{leftroot}
\restoresymbol{AMS}{uproot}

\usepackage{amssymb}
\usepackage{arxiv}
\usepackage{authblk}
\usepackage[utf8]{inputenc} 
\usepackage[T1]{fontenc}    
\usepackage[hidelinks]{hyperref}       
\usepackage{url}            
\usepackage{booktabs}       
\usepackage{amsfonts}       
\usepackage{nicefrac}       
\usepackage{microtype}      
\usepackage{lipsum}		
\usepackage{natbib}
\usepackage{doi}
\usepackage{amsmath}
\usepackage{amstext}
\usepackage{graphicx}
\usepackage{stackengine,scalerel,palatino}
\usepackage{microtype}
\usepackage{nccmath,amsmath}
\usepackage{amsfonts}
\usepackage{mathtools}
\usepackage{tensor}
\usepackage{braket}

\usepackage[nointegrals]{wasysym}

\makeatletter
\renewcommand\AB@affilsepx{\protect\\}
\makeatother
\title{Are Kronberger 80 and/or Kronberger 82 regions PeVatron candidates for LHAASO J2108+5157?}

\author[1,*]{Akash \textsc{Gupta}\thanks{email: agupta@ph1.uni-koeln.de}}
\author[2,$\dagger$]{Eduardo \textsc{de la Fuente}\thanks{Research stay at Institute of Cosmic Ray Research, University of Tokyo, 2025. Corresponding author: eduardo.delafuente@academicos.udg.mx}}
\author[3,$\ddagger$]{Ram K. \textsc{Yadav}\,\thanks{Corresponding author: ram$\_$kesh@narit.or.th}}
\author[4]{Alicia \textsc{Porras}}
\author[5]{Saurabh \textsc{Sharma}}
\author[6,7]{Sei \textsc{Kato}}
\author[8]{Daniel \textsc{Tafoya}}
\author[9]{Miguel A. \textsc{Trinidad}}
\author[10]{Alberto \textsc{Nigoche-Netro}}
\author[9]{Harold E. \textsc{Viveros}}
\author[7]{Kazumasa \textsc{Kawata}}
\author[11]{Hiromasa \textsc{Suzuki}}
\author[7]{Munehiro \textsc{Ohnishi}}  
\author[12]{Ivan \textsc{Toledano-Juarez}}
\author[7]{Takashi \textsc{Sako}}
\author[7]{Masato \textsc{Takita}}

\affil[1]{I. Physikalisches Institut, Universität zu Köln, Zülpicher Str. 77, 50937 Köln, Germany}

\affil[2]{Departamento de F\'{i}sica, CUCEI, Universidad de Guadalajara, Blvd. Marcelino Garc\'{i}a Barragan 1420, Ol\'{i}mpica, 44430, Guadalajara, Jalisco, M\'exico}

\affil[3]{National Astronomical Research Institute of Thailand (Public Organization), Chiangmai, Thailand}

\affil[4]{Instituto Nacional de Astrof\'isica, \'Optica y Electr\'onica,  Luis E. Erro N\'um.~1, Tonanzintla, Puebla, M\'exico.}

\affil[5]{Aryabhatta Research Institute of Observational Sciences, Manora Peak, Nainital 263001, India}

\affil[6]{Institute for Cosmic Ray Research, University of Tokyo, Kashiwa 277-8582, Japan}

\affil[7]{Institut d’Astrophysique de Paris, CNRS UMR 7095, Sorbonne Université, 98 bis bd Arago 75014, Paris, France}

\affil[8]{Department of Space, Earth, and Enviroment, Chalmers University of Technology, Onsala Space Observatory, 439 92, Sweden}

\affil[9]{Departamento de Astronom\'{i}a, Universidad de Guanajuato, Apartado Postal 144, 36000, Guanajuato, Guanajuato, M\'exico}

\affil[10]{Instituto de Astronom\'{i}a y Meteorolog\'{i}a, CUCEI, Universidad de Guadalajara, Av. Vallarta 2602, Guadalajara, Jalisco, M\'exico}

\affil[11]{Institute of Space and Astronautical Science, Japan Aerospace Exploration Agency, Kanagawa 252-5210, Japan}

\affil[12]{Doctorado en Ciencias en F\'{i}sica, CUCEI, Universidad de Guadalajara, Jalisco, M\'exico} 

\renewcommand{\shorttitle}{\textit{arXiv} Template}

\hypersetup{
pdftitle={A template for the arxiv style},
pdfsubject={q-bio.NC, q-bio.QM},
pdfauthor={David S.~Hippocampus, Elias D.~Striatum},
pdfkeywords={First keyword, Second keyword, More},
}

\begin{document}
\maketitle

\begin{abstract}
High-energy gamma rays have been detected in the region of LHAASO~J2108+5157 by the Fermi--LAT, HAWC and LHAASO-KM2A observatories. Cygnus~OB2 in Cygnus--X has been confirmed as the first strong stellar cluster PeVatron in our Galaxy. Thus, the star--forming regions Kronberger~80 and Kronberger~82, located in the field of LHAASO~J2108+5157, are analyzed to evaluate their stellar population and potential as associated PeVatron candidates. A distance of 10~kpc is adopted for Kronberger~80, while $\sim$1.6~kpc is estimated for Kronberger~82. Based on stellar densities, we report that their cluster radii are 2.5$\arcmin$ and 2.0$\arcmin$, while IR photometry reveals poor stellar content in massive O-type stars in both cases. From optical data, the estimation of cluster ages are 5--12.6~Myr and $\lesssim$ 5~Myr, respectively. We conclude that, in contrast to the stellar content of Cygnus~OB2, it is unlikely that Kronberger~80 and Kronberger~82 are PeVatrons associated with LHAASO~J2108+5157. The presence of a PeVatron in this region remains a mystery, but we confirm that the two Kronberger regions are star--forming regions undergoing formation rather than evolution. 
\end{abstract}

\keywords{ISM: clouds \and ISM: molecules \and ISM: individual objects (LHAASO J2108+5157, Kronberger 80, Kronberger 82) \and gamma-rays}

\section{Introduction}
\label{sec:introduction}

\begin{figure*}[!ht]
\begin{center}
\includegraphics[width=\textwidth]{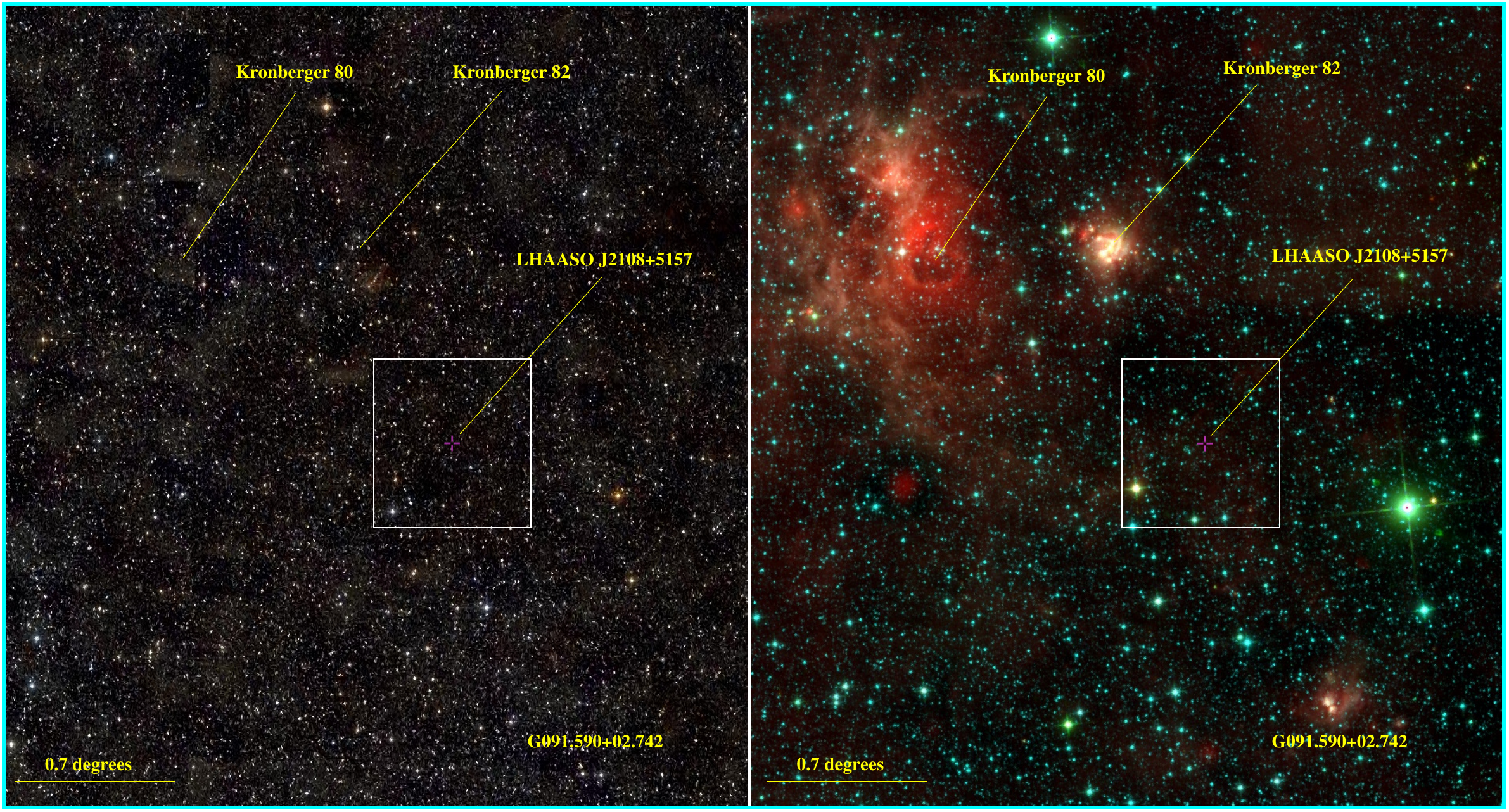}
\end{center}
\caption{Three square degrees infrared images centered at $\alpha$(J2000) = 21$^h$08$^m$58$^s$ and $\delta$(J2000) = +52$^{\circ}$00$'$03$''$ showing the field of view around LHAASO J2108+5157 located at $\alpha$(J2000) = 21$^h$08$^m$52.8$^s$ and $\delta$(J2000) = +51$^{\circ}$57$'$00$''$. The typical 2MASS image (RGB at J, H, K$_{\rm s}$) is shown at left panel, and the all--WISE (combined 22$\mu$m, 12$\mu$m, 4.6$\mu$m, and 3.4 $\mu$m) at right panel. Red nebulosity reveals warm molecular gas and dust. Kronberger~80 and 82 regions are labeled, and the white square with a size of 0.7$^{\circ}$ shows the LHAASO gamma-ray emission region. North to the top and East to the left. \\
Alt text: Two color composite images showing the portion of the sky containing the gamma-ray source LHASSO J2108+5157. One has  colors in near-infrared filters, and the other, in mid-infrared, includes location labels of the most relevant sources.}
\label{fig:Krons_IR}
\end{figure*}

PeVatrons are natural particle accelerators capable of achieving energies of up to peta-electronvolts (PeV=10$^{15}$~eV). However, the current understanding of PeVatrons is still evolving, especially concerning their origin and the nature of their leptonic or hadronic emissions. In the hadronic context, reliable radio observations of molecular and neutral gas are crucial to test this emission because they determine the density of nucleons n(H$_2$+HI)=2n(H$_2$)+n(HI), an essential parameter related to the production of gamma rays \citep{Kelner2006}. These gamma rays originate from the decay of neutral pions into two gamma rays \citep{Dermer1986} when cosmic rays (accelerated in the PeVatron) interact with molecular clouds \citep[e.g.,][]{delaFuente2023c}. 

PeVatrons are pivotal in unraveling the mysteries of the nature and origin of cosmic rays. The search for PeVatrons is guided by predictions and observational data from various facilities, including the High Energy Stereoscopic System (H.E.S.S.) in Namibia \citep[e.g.][]{Ohm2023}, the High Altitude Water Cherenkov Observatory or HAWC in Mexico \citep[e.g.][]{Abeysekara2023}, the Tibet Air Shower--$\gamma$ (Tibet~AS--$\gamma$) in Tibet \citep[e.g.][]{Amenomori2019} and the Large High Altitude Air Shower Observatory--KM2A (LHAASO--KM2A) in China \citep[e.g.][]{Ma2022}. These observatories have contributed significantly to our understanding of PeVatrons, including pioneering work \citep[e.g.][]{Abramowski2016, Abeysekara2020, Abeysekara2021, Amenomori2021a, Amenomori2021b, Amenomori2021c, Cao2021a, Cao2024}. In 2021, the LHAASO--KM2A collaboration identified 12 PeV gamma-ray galactic sources \citep{Cao2021a}. 
However, in their first catalog in 2023, the number of sources increased dramatically to 90 sources, each with a significance of 5 $\sigma$ and an extended size of $\leq$ 2$^{\circ}$ \citep{Cao2024}. Reviews on PeVatrons have been published by \citet{Anguner2023, Cao2023, Cristofari2021, delaFuente2023a, Owen2023} and references therein.

The association of massive star-forming regions and PeVatrons is supported by the confirmation of Cygnus OB2 (hereafter Cyg-OB2) as the first massive star cluster identified as a PeVatron by HAWC and LHAASO–KM2A \citep[e.g][and references therein]{Abeysekara2021,Cao2021a,Cao2024}. Cyg-OB2 is associated with the Cygnus cocoon region (Ackermann et al. 2011), located in the Cygnus-X molecular cloud \citep[e.g.][]{Reipurth2008,Schneider2006}.
Cyg-OB2 is one of the most massive OB associations in the Galaxy, with nearly hundreds of OB stars, an age between 1 and 7 Myrs, and a strong stellar wind power of $\sim$10$^{39}$~erg~s$^{-1}$ \citep [e.g.][and references therein]{Knodlseder2000,Lozinskaya2002}. Recently, Cao et al. proposed Cyg-OB2 as a new type of object, a PeVatron that accelerates protons to at least 10 PeV: a \textit{super-PeVatron} \citep{LHAASO2024}. The PeVatron candidate LHAASO J2108+5157 (hereafter J2108) is considered one of the most enigmatic in the list, as no counterpart has been identified since its discovery. This counterpart remains a mystery, but due to its location in the optical images, \citet{Cao2021a,Cao2021b} originally proposed the star-forming regions Kronberger~80 (hereafter Kron~80) and Kronberger~82 (hereafter Kron~82) as PeVatron candidates. However, these authors discarded Kron~80 as PeVatron candidate due to its far distance (5.5~kpc). 

In this work, we analyze the stellar content in Kron~80 and Kron~82, searching for an association with J2108 as the PeVatron optical counterpart. Our work is pioneering for Kron~80, and the first for Kron~82. In the latter case, we derive a more realistic distance (\S~\ref{sec:distanceKron82}). We quantify for the first time the stellar content of J2108, in the reported LHAASO emission region of 0.7$^\circ$ diameter, according to \citet[][Fig.~2]{delaFuente2023b} by searching for stellar clusters and young stellar objects (YSO) with IR photometry. 

Section~\ref{sec:overview} gives a brief overview of J2108 and the two Kronberger sources. The photometric methods and observations used in this study are described in section~\ref{sec:obs}. The results and discussions are presented in Section~\ref{sec:res_disc}, while the conclusions are presented in Section~\ref{sec:conclusion}. To complement the optical and IR study, and in the case of the null hypothesis that Kron~80 and Kron~82 are not the PeVatrons associated with J2108, we speculate in Appendix~\ref{appendix:ap1} on the density of nucleons required in the vicinity of J2108 to produce the observed (sub)PeV emission, given the closest PeVatron known to date.

\section{Sources Overview}
\label{sec:overview}

\subsection{LHAASO J2108+5157 in Cygnus OB7}
\label{sec:J2108}

The first detailed study of J2108 located at {\it l} = 92.30$^{\circ}$ and {\it b} = 2.84$^{\circ}$ in the molecular cloud of Cygnus OB7 (hereafter Cyg-OB7) was presented in \citet{Cao2021b}. Although these authors assumed a distance of 3.3~kpc, Cyg-OB7 is related to Cygnus-X molecular cloud with a range of distances between 600 pc and 1.7 kpc \citep[e.g.][and references therein] {Schneider2006}.

\citet{Cao2024}, introduced the term \textit{dark source} to describe gamma-ray sources without associations or counterparts at other wavelengths. In that catalog, J2108 is listed as LHAASO J2108+5153u\footnote{The `u' in the designation stands for ultra-high energy sources with a test statistic at 100 TeV (TS$_{\rm 100}$) of more than 20.}. However, it was not classified as a dark source, further emphasizing its mysterious properties. Comprehensive reviews, multi-wavelength studies, and detailed analyses can be found in \citet{DeSarkar2024, Mitchell2024, Abe2023, delaFuente2023b, Cao2021b} and the references therein. No photometric studies are reported for this region, and specific studies to explain gamma-ray emission associated with supernova explosions are discussed in \citet{DeSarkar2024, Mitchell2024, DeSarkar2023, Kar2022}.

J2108 was associated by \citet{Cao2021b} with the molecular cloud MML[2017]4607 (hereafter MML) by low-resolution ($\sim$8.5$'$) optically thick observations at $^{12}$CO(J=1$\rightarrow$0) with a 1.2m radio telescope \citep{Miville2017,Dame2001}. Further studies with higher resolution using optically thin observations of $^{13}$CO(J=2--1) and $^{13}$CO(J=1--0) were performed with two radio telescopes: the 1.85-meter Osaka Prefecture College Radio Telescope (OPURT), which provides a resolution of 2.7$'$, and the 45-meter Nobeyama Radio Telescope (NRO45m), which provides a resolution of 17$''$. These studies, reported by \citet{delaFuente2023b,delaFuente2023c}, show that J2108 is not associated with only one cloud, possibly with two molecular clouds. These are called [FKT-MC] 2022 \citep[][; hereafter FKT2022]{delaFuente2023b} and [FTK-MC] \citep{delaFuente2023c}. In particular, MML is located in the [FTK–MC] molecular cloud, which has a larger extension than MML. Although the distance proposed to MML of 3.3 kpc was determined considering the rotation curve of \citet{Brand1993}, the distances reported of 1.63 and 1.70 kpc for [FTK-MC] and FKT2022 were determined using a more reliable rotation curve \citep{2019ApJ...885..131R} as described in \citet{delaFuente2023b,delaFuente2023c}. Considering the Cyg-OB7 distance range, the distance of 3.3 kpc is not realistic, so it is worth investigating MML at the closest distance ($\sim$ 1.6 kpc) and with optically thin observations.

\subsection{Kronberger 80 and Kronberger 82}
\label{sec:krons}

Kron~80 (DSH J2111.8+5222) with $\alpha_{2000}$=21$^{\rm h}$11$^{\rm m}$50.5$^{\rm s}$; $\delta_{2000}$=+52$^{\circ}$22$'$48$''$, and Kron~82 (DSH J2109.5+5223) with $\alpha_{2000}$=21$^{\rm h}$09$^{\rm m}$30.4$^{\rm s}$; $\delta_{2000}$=+52$^{\circ}$23$'$41$''$ were catalogued by \citet{Kronberger2006} as candidates for open star clusters. These authors report visual diameters of 0.9$'$ and 4.0$'$, respectively. Both regions have been poorly studied.

For Kron~80, \citet{Kharchenko2012} reported a photometric distance of 5 kpc with a color excess E(B-V) = 1.29 mag, while \citet{Cantat-Gaudin2018} (using {\it Gaia} DR2) reported a galactocentric distance of 13.36 kpc and 41-star members. \citet{Molina-Lera2019} reported detailed photometry in the \textit{ugri} bands giving an age between 10--20~Myr and a radius of 2.3$'$. Due to the distance of 5 kpc, \citet{Cao2021b} discarded it as a PeVatron candidate for J2108, but no quantitative argument was provided.

Kron~82 refered to the Northern Coalsack \citep[NCS, ][]{Dobashi2014,Dobashi2018}, was summarized and studied in detail by \citet[][and references therein]{delaFuente2023b,Dobashi2018}. \citet{delaFuente2023b} reported OPURT $^{12,13}$CO(J=2$\rightarrow$1) observations assuming a distance range between 1.6 kpc \citep[from][]{Moscadelli2021} and 2.3 kpc (far kinematic distance), and \citet{Dobashi2014,Dobashi2018} reported NRO45m observations at $^{12,13}$CO(1$\rightarrow$0) and C$^{18}$O(1$\rightarrow$0) assuming a distance of 800~pc \citep{Humphreys1978} and T$_{\rm exc}$ of 52~K for a component velocity centered in --6~km~s$^{-1}$. \citet{Dobashi2018} from C$^{18}$O reported a M(H$_2$)~$\sim$~1$\times$10$^{3} M_{\odot}$, typical of clumps with young clustering \citep[][and references therein]{Shimoikura2018}, and showed that V$_{\rm LSR}$ associated with the turbulent Cyg-OB7 gas is --4~km~s$^{-1}$. \citet{delaFuente2023b} reported physical parameters considering a T$_{\rm exc}$ of 16 K obtaining a $^{13}$CO(2--1) H$_2$ mass between 300 and 400~M$_{\odot}$. There are no photometric studies of Kron~82 in the literature, and the distance needs to be clarified. However, as shown in figure~2 of \citet{delaFuente2023b} and opposite to Kron~80, Kron~82 presents 20 cm radio-continuum NVSS emission related to ionized gas. 

In Fig.~\ref{fig:Krons_IR} we show near- and mid-infrared images of Kron~80 and Kron~82 taken with the observatories 2MASS \citep[classical RGB, left panel;][]{Skrutskie2006} and WISE \citep[at 3.4$\mu$m, 4.6$\mu$m, 12$\mu$m and 22$\mu$m, right panel;][]{Wright2010} at the same scale and including the J2108 region. In contrast to J2108, we observe weak extended emission associated with Kron~80 and Kron~82 in 2MASS, which is more prominent in all-WISE bands. Due to the lack of \textit{Spitzer}--IRAC images at 5.8$\mu$m and 8.0$\mu$m, as well as MIPS 24.0$\mu$m and the fact that the images at 3.6$\mu$m and 4.5$\mu$m do not cover the eastern part of Kron~80, we performed our NIR-MIR analysis using only the 2MASS and WISE bands. However, and despite the lack of IRAC and MIPS observations, it is clear that (embedded) star clusters, gas and dust coexist in Kron~80 and Kron~82, as discussed in \citet[][]{delaFuente2023b,delaFuente2020a,delaFuente2020b,delaFuente2009,Churchwell2007} and references therein, for star-forming regions.

\section{Observations and Methods}
\label{sec:obs}

\subsection{Optical and infrared observations}
\label{subsec:opt-nir}

Analysis of optical and IR wavelength observations is conducted to i) describe the stellar content associated with J2108, Kron~80 and Kron~82 regions, ii) identify if (proto-)stellar spatial distributions are clustered, and iii) quantify if the number of intermediate-high mass stars interacting with the molecular material is enough to justify the possible nature of these regions as PeVatrons for the observed gamma-ray emission.

We use the deep photometric data from the IPHAS \citep[INT Photometric H$\alpha$ Survey of the Northern Galactic Plane DR2, ][]{2014MNRAS.444.3230B} to analyze stellar content in J2108, Kron~80 and Kron~82 and to estimate cluster ages (\S~\ref{sec:ages}).

For NIR photometry, we retrieve data for J2108, Kron~80, and Kron~82 from the Two--Mircron All Sky Survey \citep[2MASS,][]{Skrutskie2006} in all bads: J (1.25$\mu$m), H(1.65$\mu$m) and K$_{\rm s}$ (2.17$\mu$m) bands, with an angular resolution of 4$''$ and using this error criterion on the filters:

\begin{equation}
 \sigma (J),\,\sigma (H),\,\sigma (K) < 0.15\,{\rm mag}.
\end{equation}

We use these data to estimate the cluster centers and cluster radii for Kron~80 and Kron~82 (see \S~\ref{sec:cluster_analysis}).
For the respective color-color (CC) and color-magnitude (CM) diagrams, we first perform color transformations to the \citet{Bessell1988} system (see \S~\ref{sec:J2108_phot}).

For MIR photometry, we use data from the Wide--field Infrared Survey Explorer \citep[WISE,][]{Wright2010} on the bands W1 (3.4$\mu$m), W2(4.6$\mu$m), W3(12$\mu$m) and W4 (22$\mu$m) with an angular resolution of 6.1$''$, 6.4$''$, 6.5$''$ and 12.0$''$, respectively. CC diagrams are performed following \citet{Kang2017} and presented in \S~\ref{sec:J2108_phot}.

\subsection{Cluster Analysis}
\label{sec:cluster_analysis}

Initial inspection of the J2108 region using optical and NIR wavelengths revealed no evident star clustering. Consequently, an iso-density analysis was deemed unsuitable for this region. Instead, our study focuses on clustering in the Kronberger regions. For the cluster analysis of Kronberger regions, we selected a common field region without nebulosity, located at $l$=92.4007200$^\circ$ and $b$=3.3503535$^\circ$, for both Kron~80 and Kron~82.

\subsubsection{Center Estimation}

The centers of the open clusters Kron~80 and Kron~82 were determined using a Kernel Density Estimator (KDE) applied to 2MASS photometric data. Stars with photometric errors in the $K$-band of less than 0.15 mag were selected for this analysis. The KDE method is well-suited for identifying the peak density regions in a stellar distribution, as it accounts for the underlying spatial distribution without assuming a predefined binning structure.

\begin{figure}[h!]
  \centering
  \begin{subfigure}[b]{0.45\textwidth}
    \includegraphics[width=\textwidth]{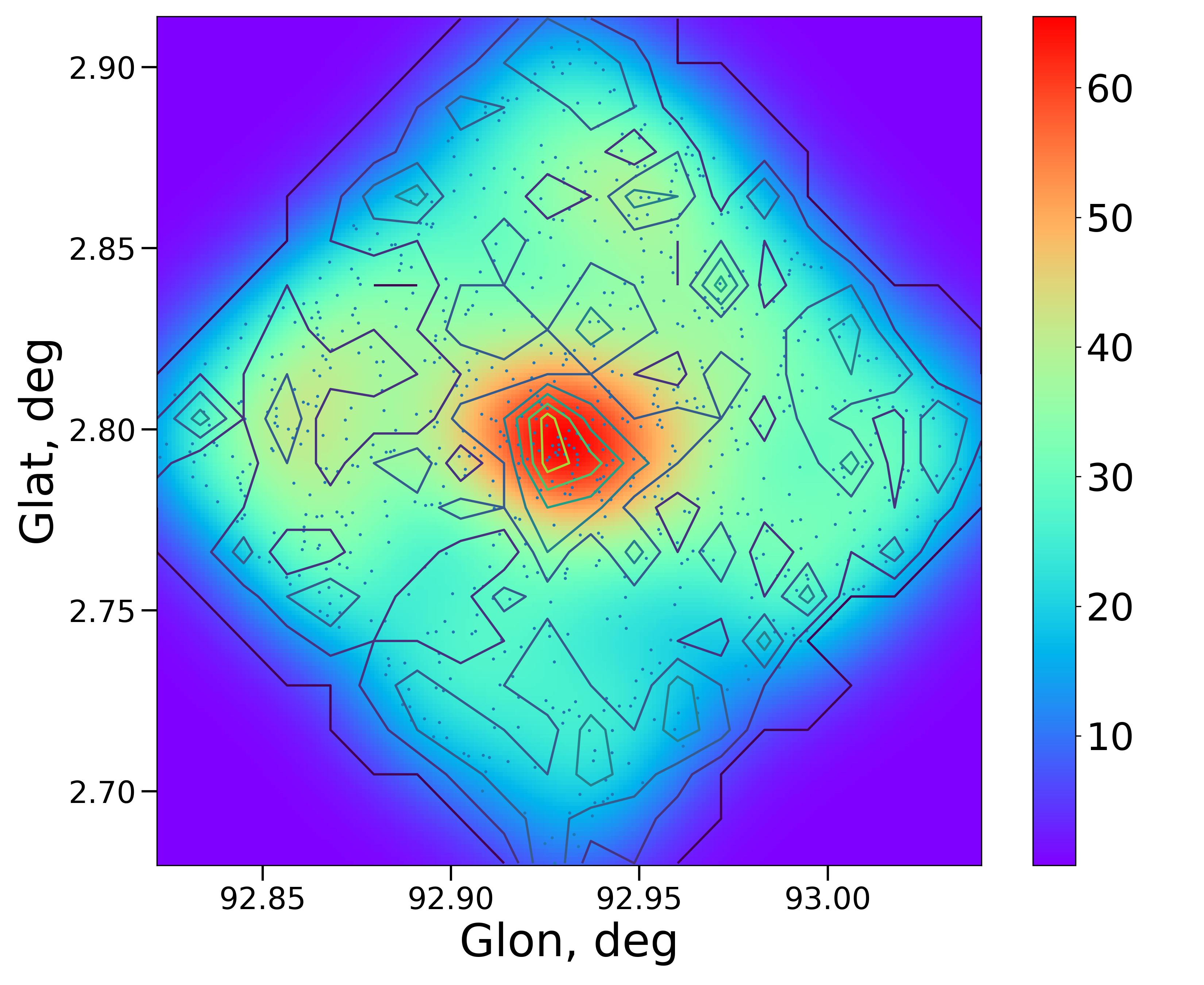}
    \label{fig:left}
  \end{subfigure}
  \hfill
  \begin{subfigure}[b]{0.45\textwidth}
    \includegraphics[width=\textwidth]{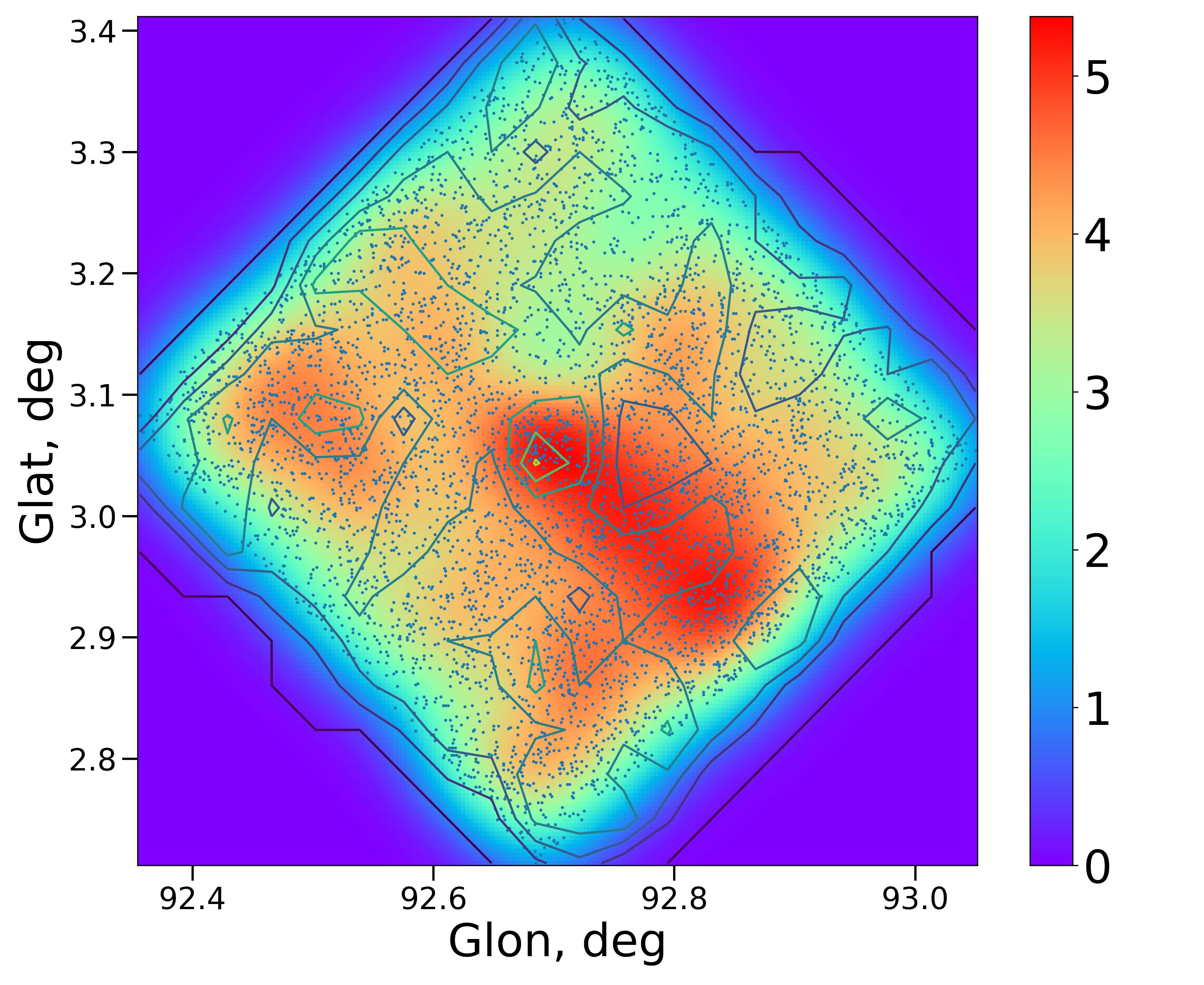}
  \end{subfigure}

\caption{Iso-density plots of Kronberger~80 (left) and Kronberger~82 (right) from 2MASS data. The color bar shows the color gradient of the Kernel Density Estimates (KDE). The scale of the 2D histogram goes from red (highest) to blue (lowest) density bins.  \\
Alt text: Color plots of isodensity contours of the Kronberger~80 and Kronberger~82 regions in galactic coordinates.}
\label{fig:iso_dens}
\end{figure}

To construct the density maps, we applied KDE to the Galactic coordinates ($l, b$) of the selected stars. The bandwidth parameter, which controls the smoothness of the KDE, was optimized separately for each cluster to ensure an accurate representation of the stellar density distribution. The resulting 2D density maps are shown in Fig.~\ref{fig:iso_dens}.

Cluster centers were identified as the coordinates of the highest density peaks in these KDE-generated maps. For Kron~80, the peak density corresponds to Galactic coordinates $l = 92.9315^{\circ}$, $b = 2.7963^{\circ}$, while for Kron~82, the peak is located at $l = 92.6879^{\circ}$, $b = 3.0468^{\circ}$.

The 2D histogram density grids were generated using a bin size of {20} for Kron~80 and Kron~82. These newly derived central coordinates serve as a reference for further analysis, including the determination of cluster radii (see below \S \ref{sec:radius_estimation}).

\subsubsection{Radius Estimation}
\label{sec:radius_estimation}

\begin{figure}[h!]
  \centering
  \begin{subfigure}[t]{0.45\textwidth}
    \includegraphics[width=\textwidth]{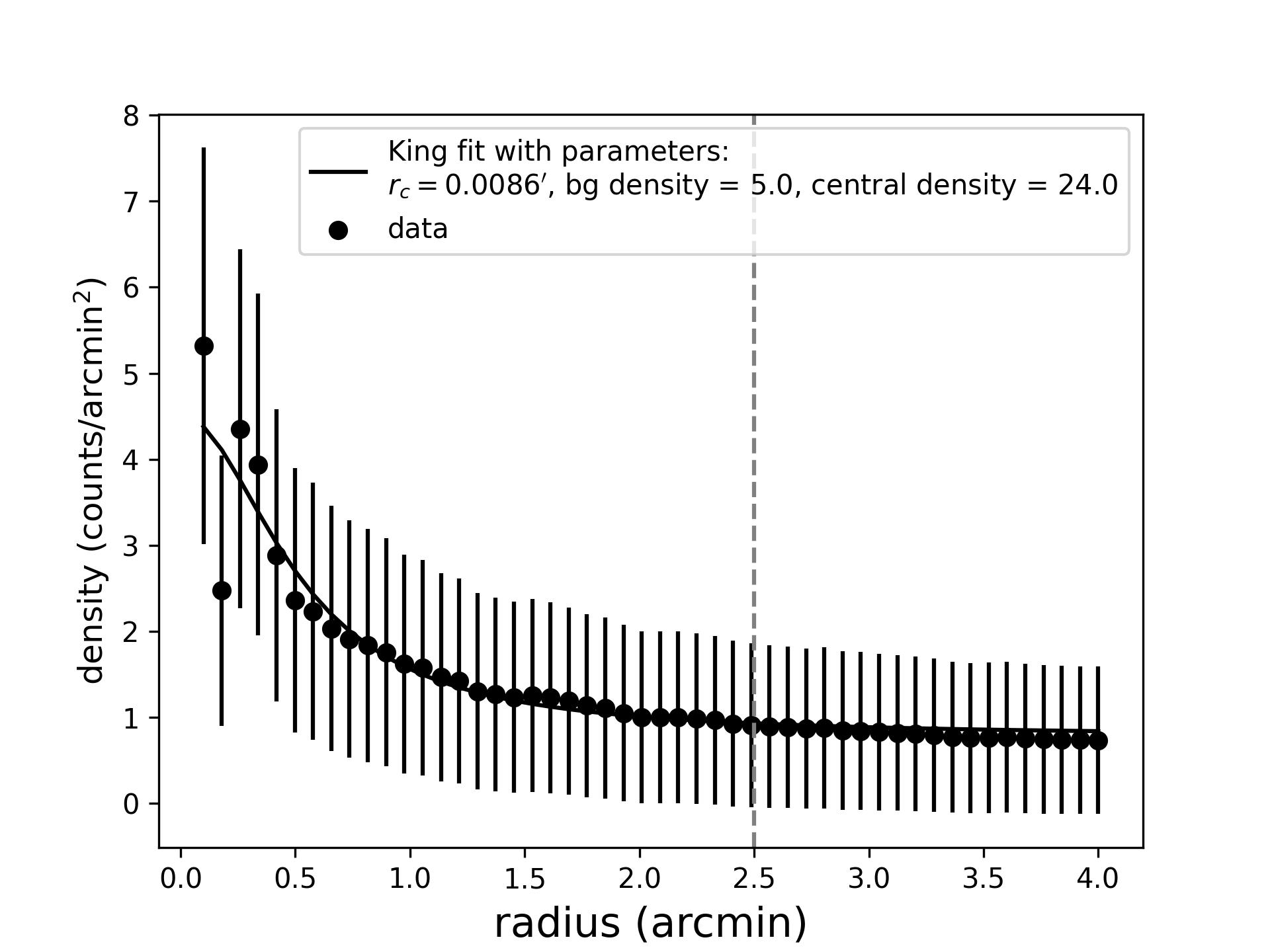}
    \label{fig:left}
  \end{subfigure}
  \hfill
  \begin{subfigure}[t]{0.45\textwidth}
    \includegraphics[width=\textwidth]{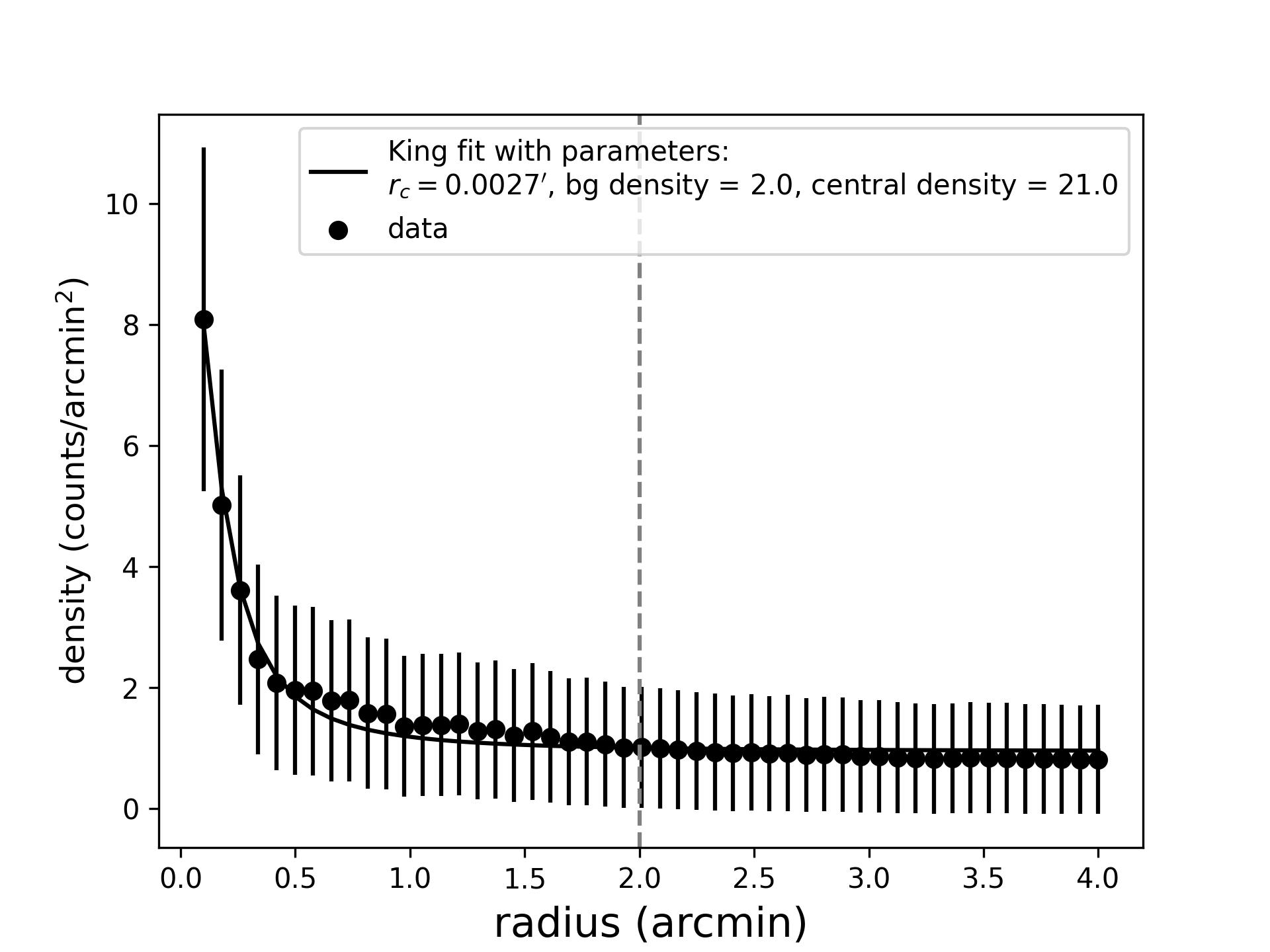}
  \end{subfigure}

\caption{King profile fitting for Kronberger~80 (left) and 82 (right). King fit parameters are summarized in the legend of these plots. Vertical dotted lines indicate the angular size of the radius obtained in each case. \\
   Alt text: Radial stellar density profiles of star clusters in Kronberger~80 and Kronberger~82. The curve of best fit for the data and the limit of the measured radii are shown.}
\label{fig:king_fit}
\end{figure}

The radial extent of a cluster is a crucial parameter for examining the properties of clusters. We derived the radial density profile (RDP) for the Kron~80 and Kron~82 cluster regions using 2MASS data to explore this. The 2MASS data were used instead of the optical data because of the higher extinction and nebulosity in these regions. We counted sources in concentric annuli, each 10 arcseconds wide, centered around the cluster's core, and normalized these counts by their respective areas. The resulting density (normalized counts) plotted against the radius is referred to as the RDP, illustrated in Fig. \ref{fig:king_fit}. 

We use King model \citep[][]{1962AJ.....67..471K} to fit the radial density profile of the clusters and to obtain their radii through the relation 

\begin{equation}
    \rho (r) = \rho_b + \frac{\rho_0}{1 + (\frac{r}{r_c})^2}~,
    \label{eq:king}
\end{equation}

\noindent where, $\rho_b, \, \rho_0, \, r_c$, denote the background stellar density, the peak density, and the core radius, respectively.
Fig.~\ref{fig:king_fit} shows the fittings to stellar density measurements around the central cluster positions. The radial density profile shows a sharper decrease from the center in Kron~82 than in Kron~80.
Based on the intersection between the observational data and the King profile given by Eq.~\ref{eq:king}, we estimate a radius of $2.5\arcmin$ and $2\arcmin$ for Kron~80 and Kron~82, respectively. For Kron~82, an angular size of $4\arcmin$ was estimated by \citet{Kronberger2006} based on visual inspection, while \citet{2020A&A...640A...1C} estimated an angular size of $1.8\arcmin$ for Kron~80 using {\it Gaia} data.

\begin{figure}[h!]
    \centering
    \begin{subfigure}[t]{0.48\textwidth}
        \includegraphics[width=\textwidth]{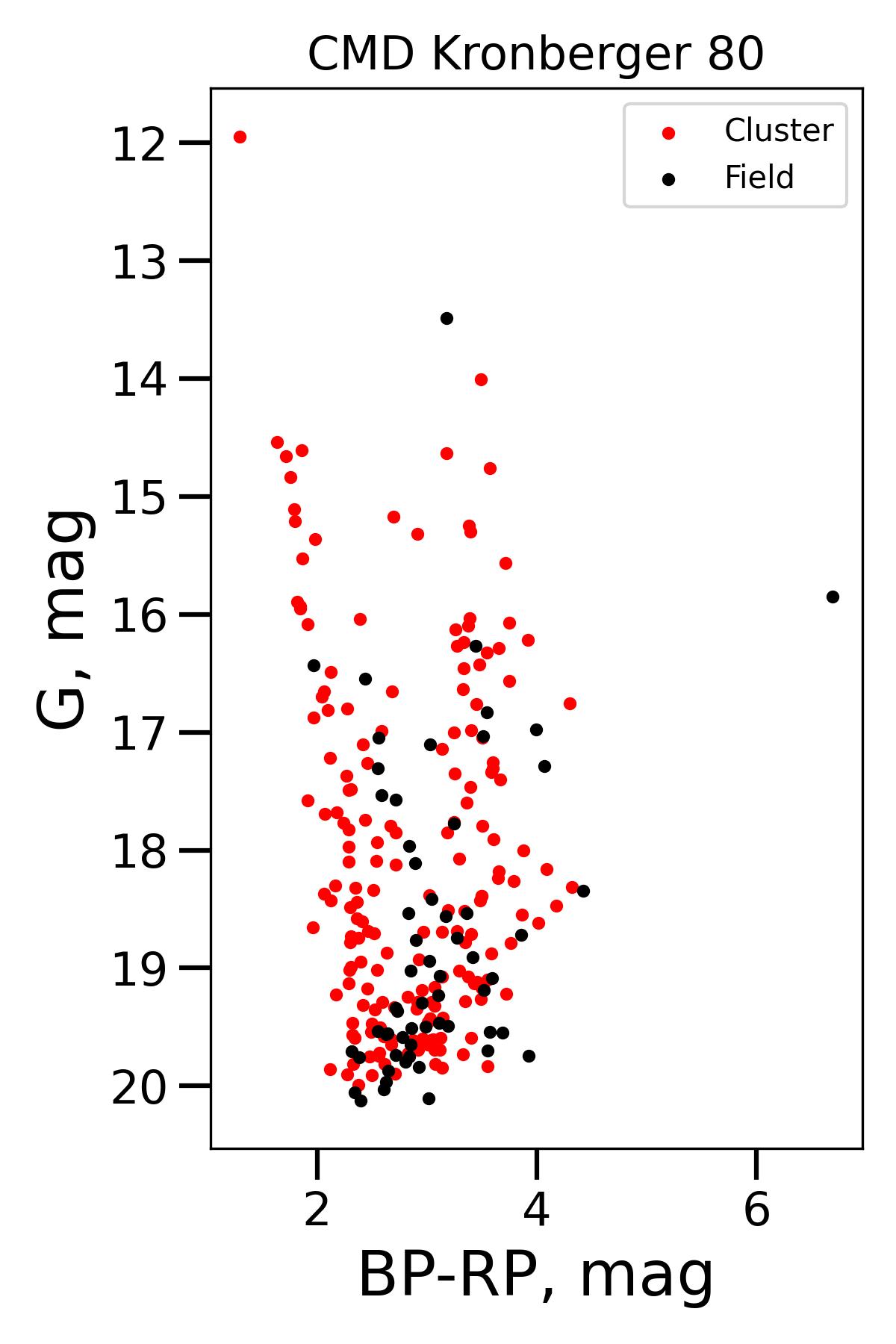}
    \end{subfigure}
    \hfill
    \begin{subfigure}[t]{0.48\textwidth}
        \includegraphics[width=\textwidth]{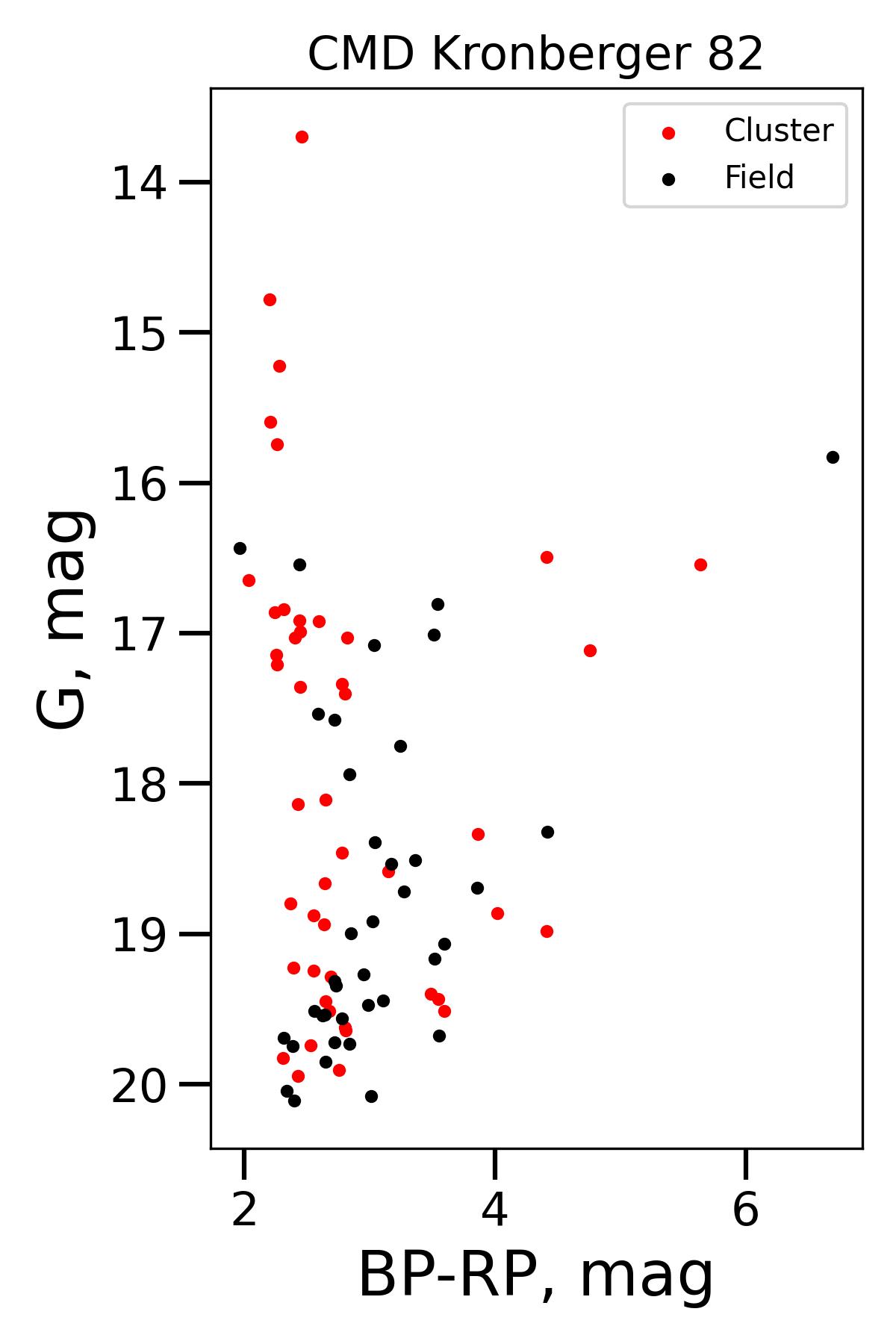}
    \end{subfigure}


    \begin{subfigure}[t]{0.48\textwidth}
        \includegraphics[width=\textwidth]{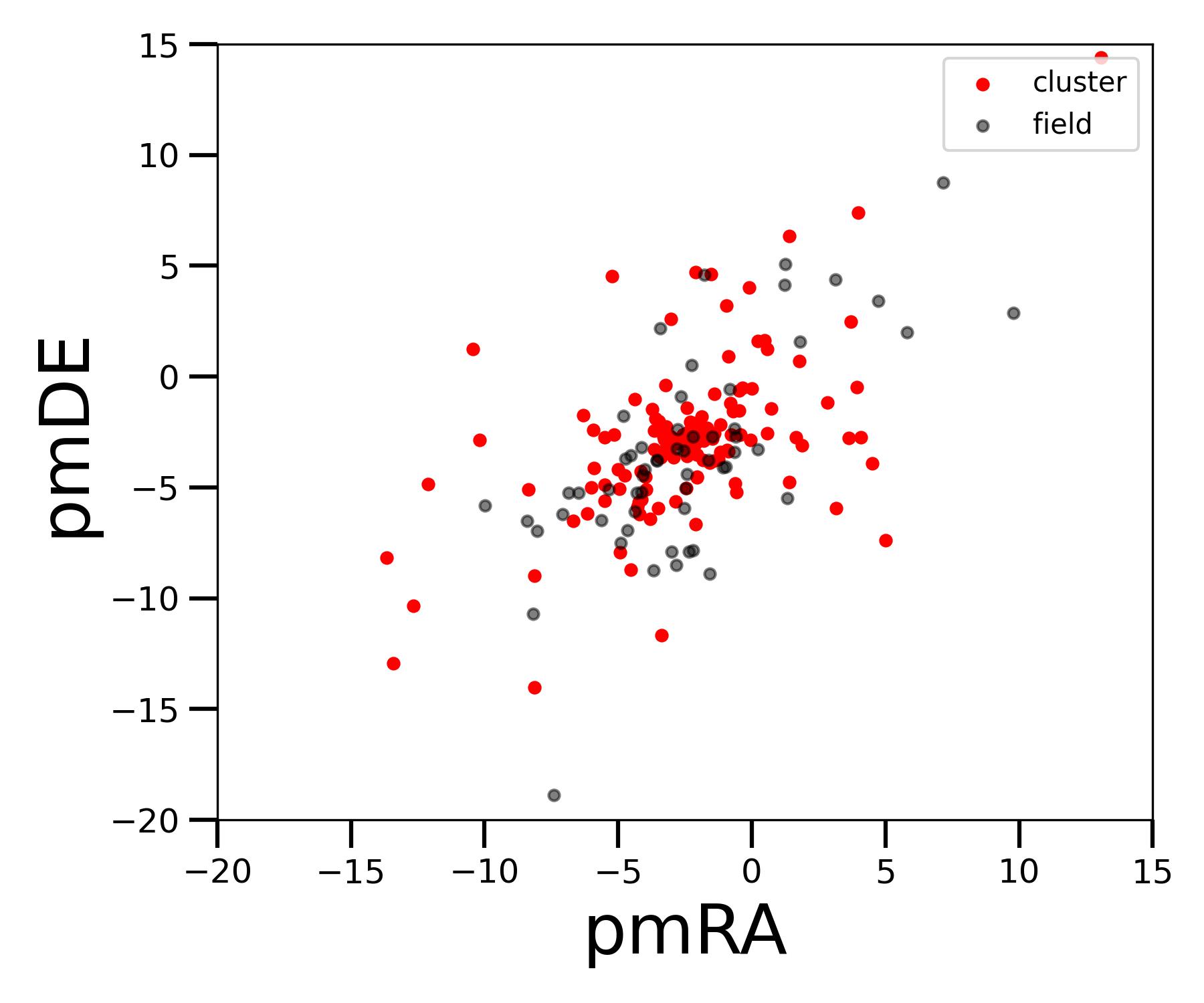}
    \end{subfigure}
    \hfill
    \begin{subfigure}[t]{0.48\textwidth}
        \includegraphics[width=\textwidth]{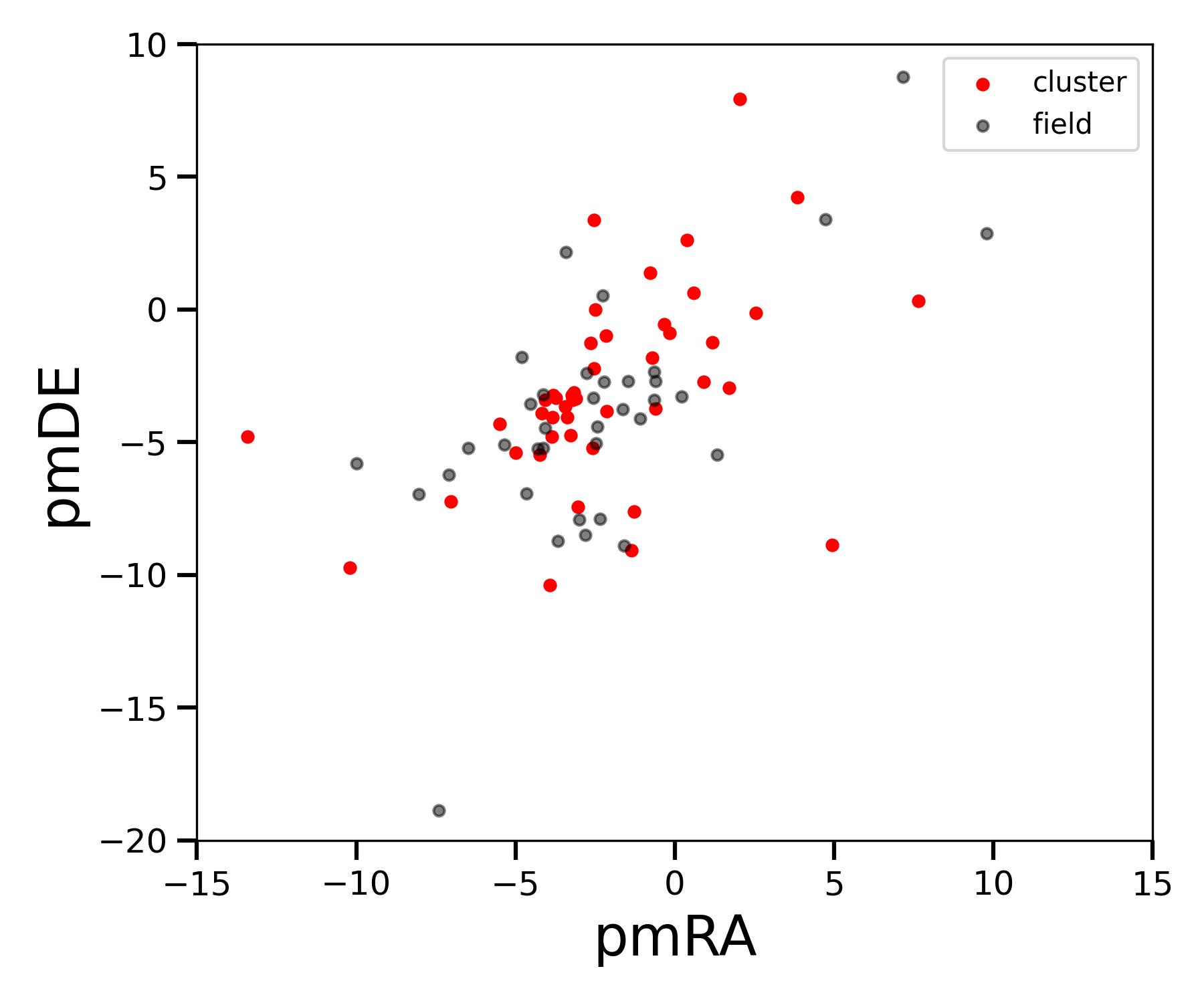}
        \caption{Kronberger 82 Vector Plot}
    \end{subfigure}

    \caption{{\bf Top}: {\it Gaia} color-magnitude diagram for Kronberger~80 (left) and Kronberger~82 (right). {\bf Bottom}: Respective vector plot diagrams. \\
    \textbf{Alt text:} Four diagrams are shown with optical photometric data obtained from the {\it Gaia} survey, for the Kronberger~80 and Kronberger~82 regions.}
    \label{fig:Gaia_cmd}
\end{figure}

\begin{figure}[htbp]
\begin{center}
\includegraphics[height=0.34\textwidth]{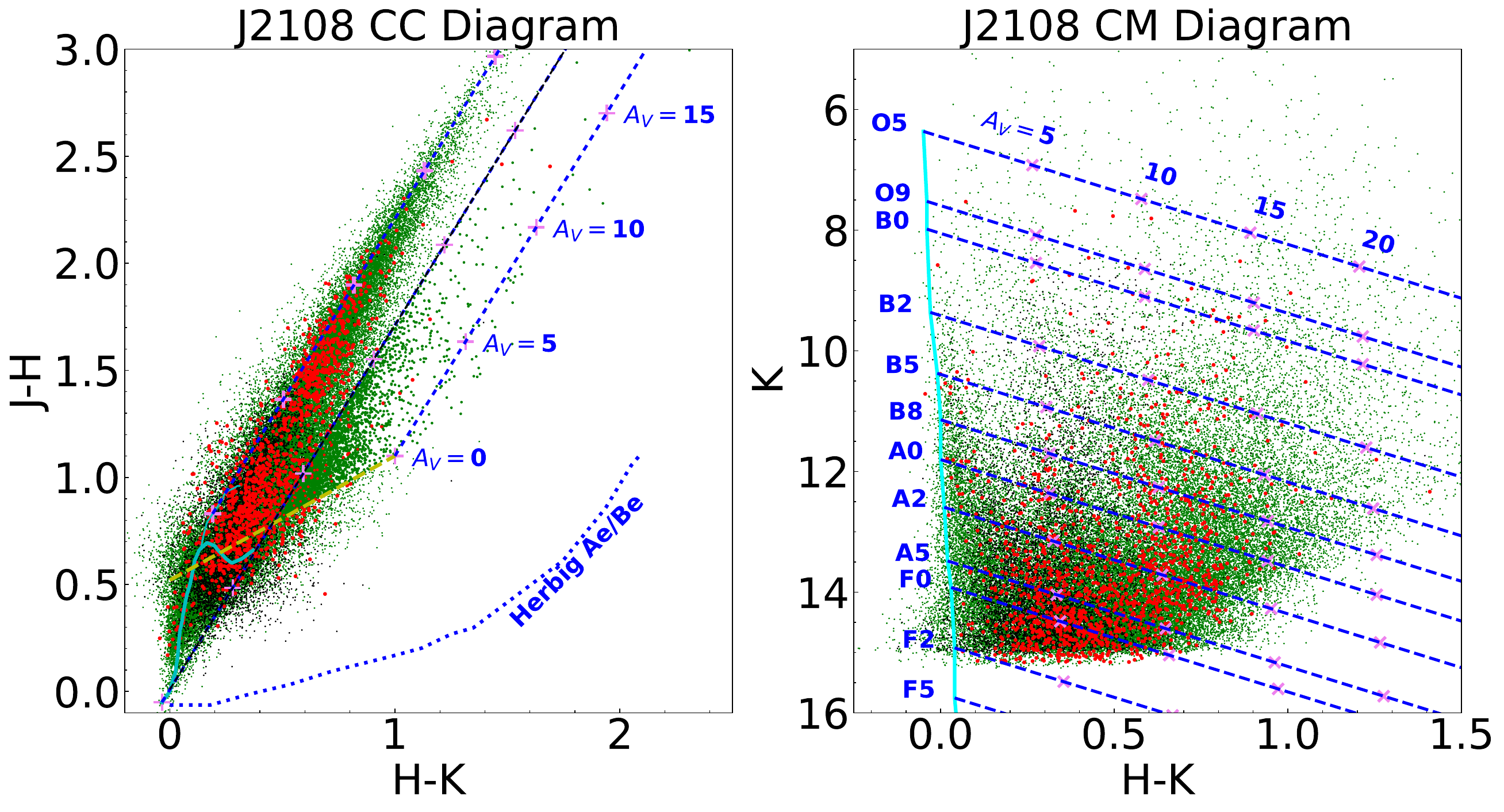} 
\includegraphics[height=0.34\textwidth]{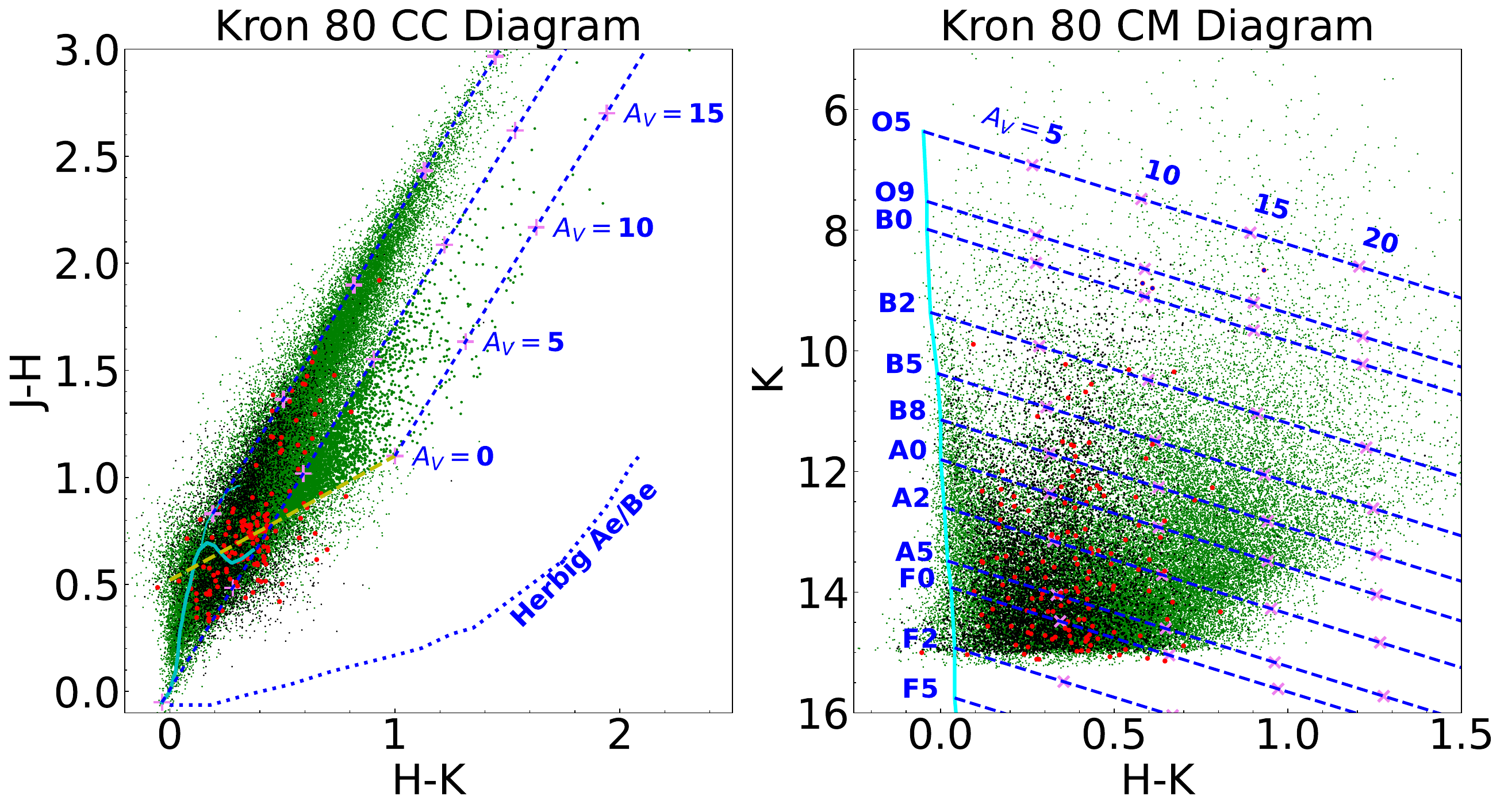}
\includegraphics[height=0.34\textwidth]{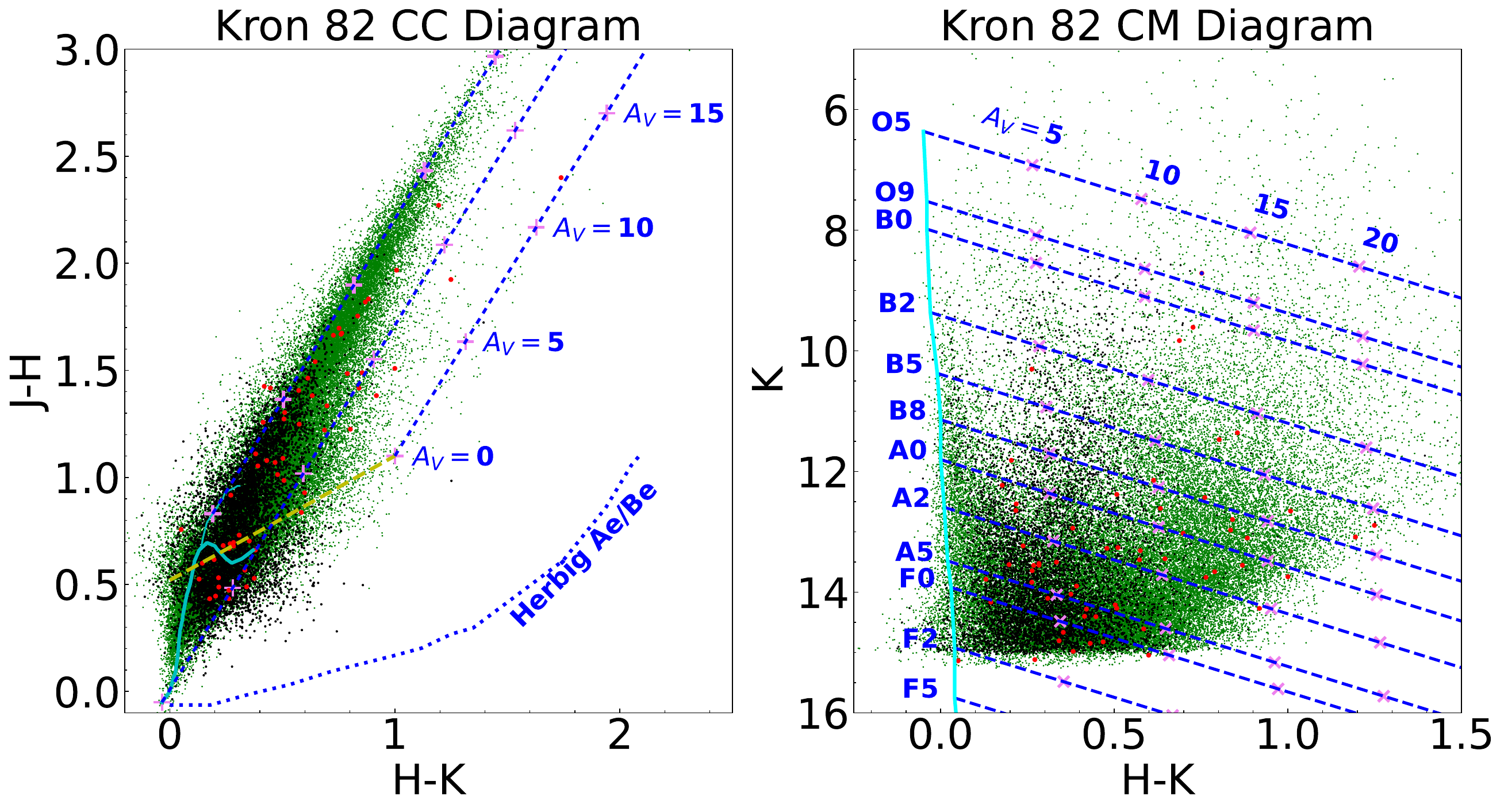}
\end{center}
\caption{
Typical 2MASS color-color (CC) and color-magnitude (CM) diagrams in \citet{Bessell1988} system for LHAASO J2108+5157 (top), Kron~80 (middle) and Kron~82 (bottom), all in red points and in comparison with Cyg-OB2 in green points. Black points are sources in a control field at 2.0$^{\circ}$ of distance from J2108, ($\alpha_{2000}$ = 20$^{h}$58$^{m}$15.1$^{s}$, $\delta_{2000}$ = 53$^{\circ}$07$'$41.5$''$, size of 2.0$^{\circ}$). \textbf{Left.} CC diagrams where the cyan lines show the location of the dwarf and giant Main Sequence \citep{Bessell1988,Koornneef1983}, the yellow dashed line represents the T-Tauri locus \citep{Meyer1997} and the blue dotted line shows the limiting position of the YSO Herbig Ae/Be objects \citep{Lada1992}. Blue dashed parallel lines represent reddening vectors according to the extinction law of~\citet{Rieke1985} with $R_{\mathrm{v}}~\approx~3.09$. Magenta crosses represent the visual extinction $A_{V}$ separated by 5 mag. \textbf{Right.} CM diagrams where the thick cyan continuum line indicates the location of the ZAMS with spectral types labeled on it \citep{Schmidt-Kaler1982}, and reddening vectors similar to the CC diagrams. \\
Alt text: Six plots of combinations of the J, H and K near-infrared filters of the LHASSO J2108+5157, Kronberger~80 and Kronberger~82 regions compared to the Cygnus OB2 region (See text for a detailed description of the stellar content).}
\label{fig:J2108_2MASS}
\end{figure}

\begin{figure*}[h!]
\begin{center}    
\includegraphics[scale=0.18]{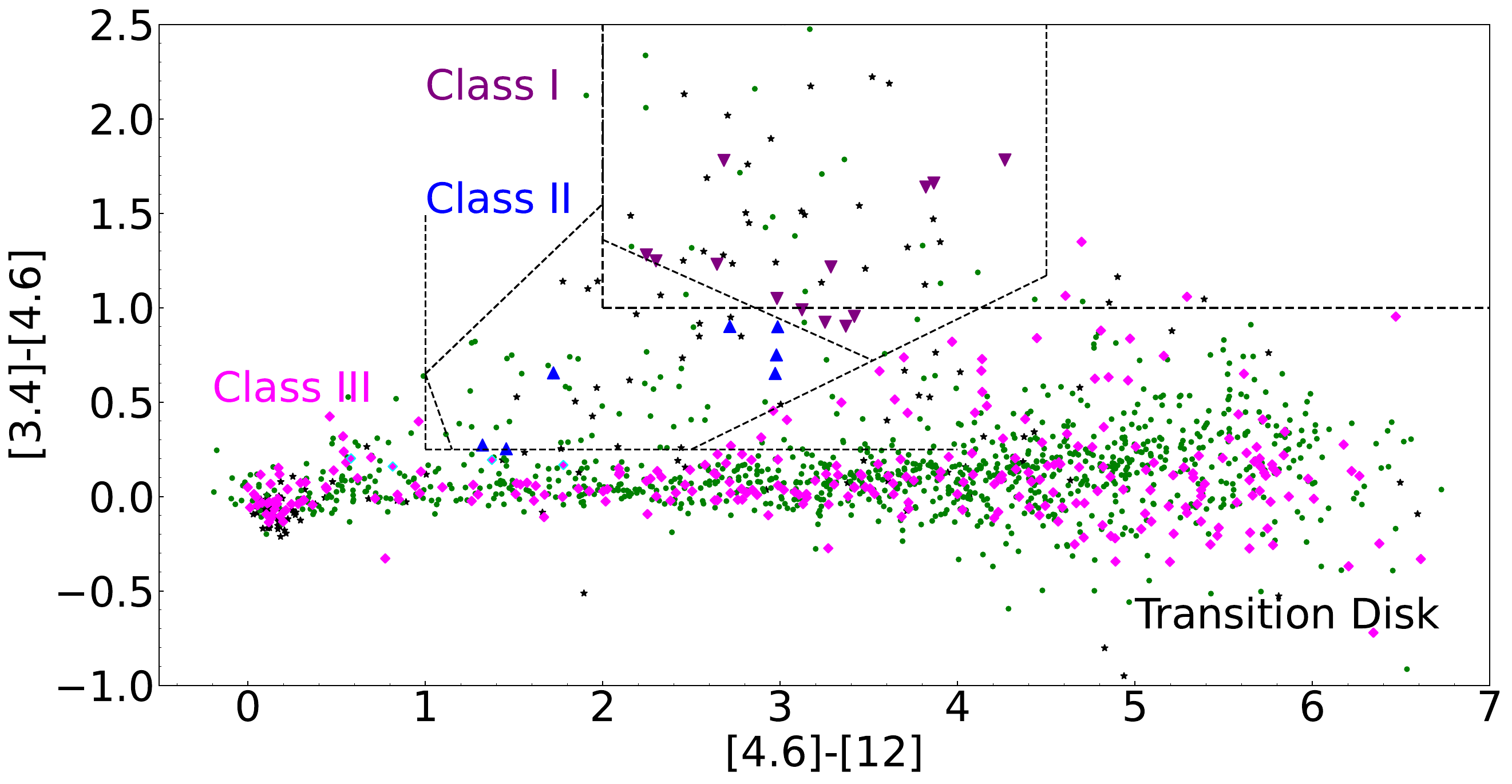}
\includegraphics[scale=0.18]{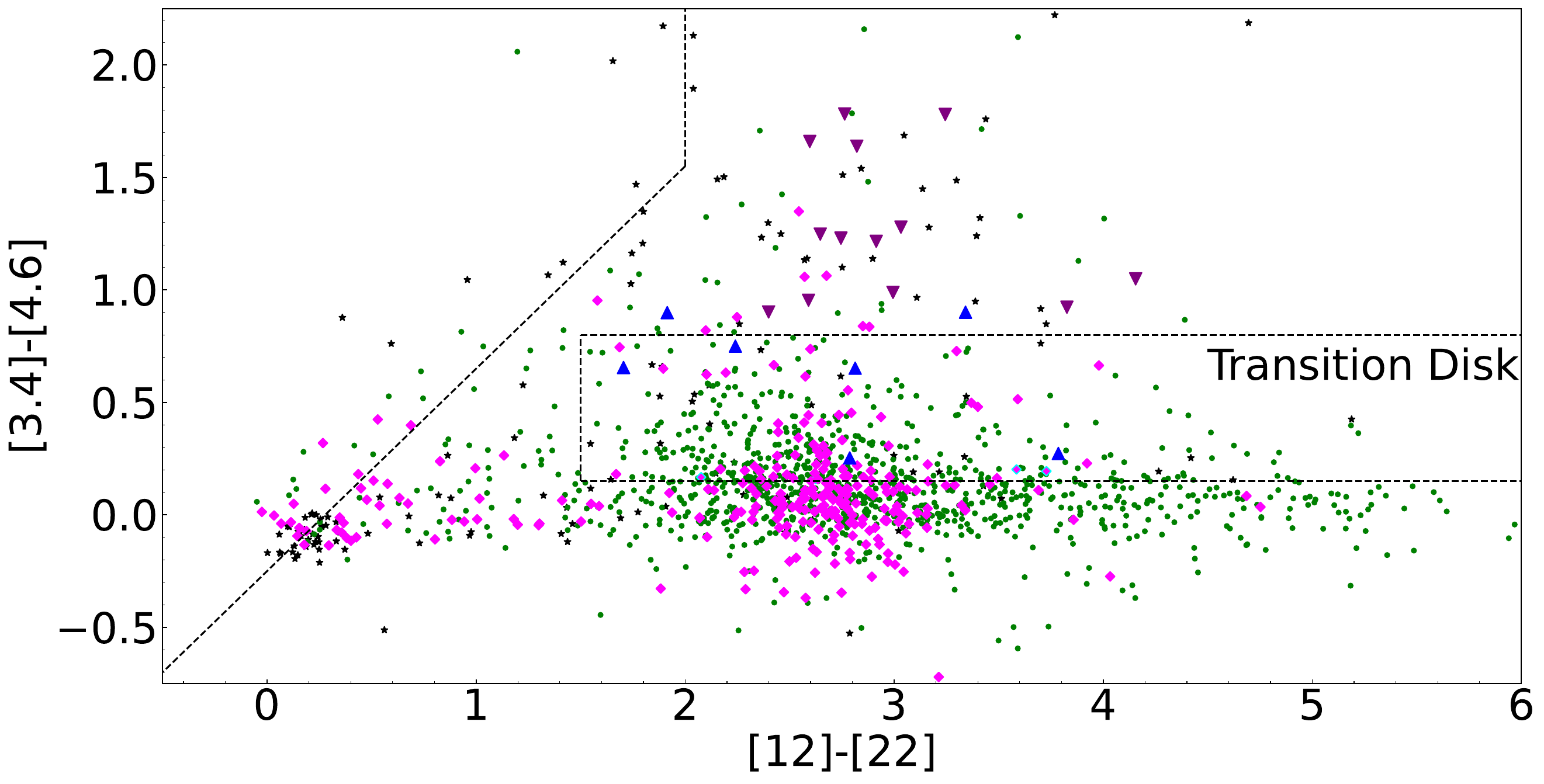} \\
\includegraphics[scale=0.18]{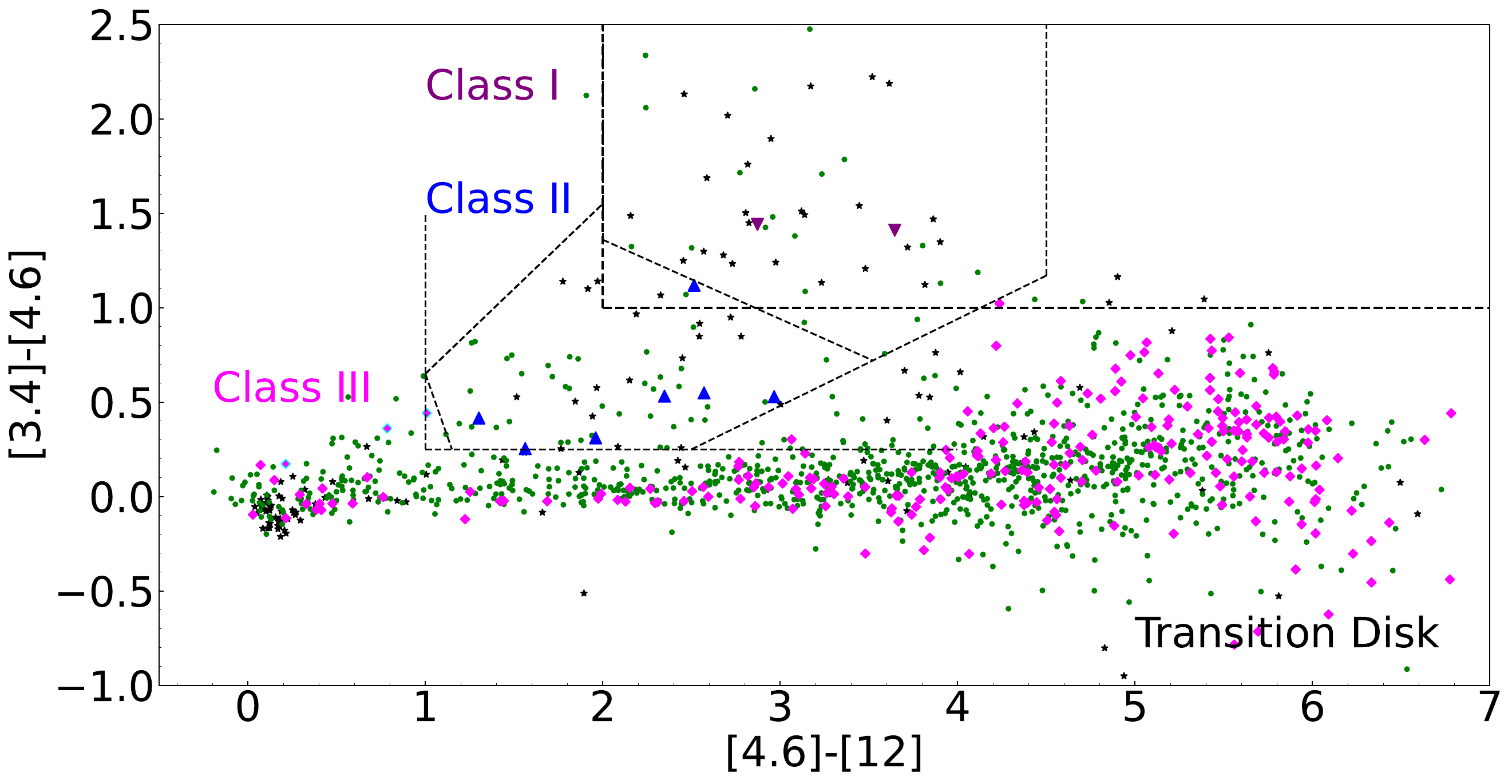}
\includegraphics[scale=0.18]{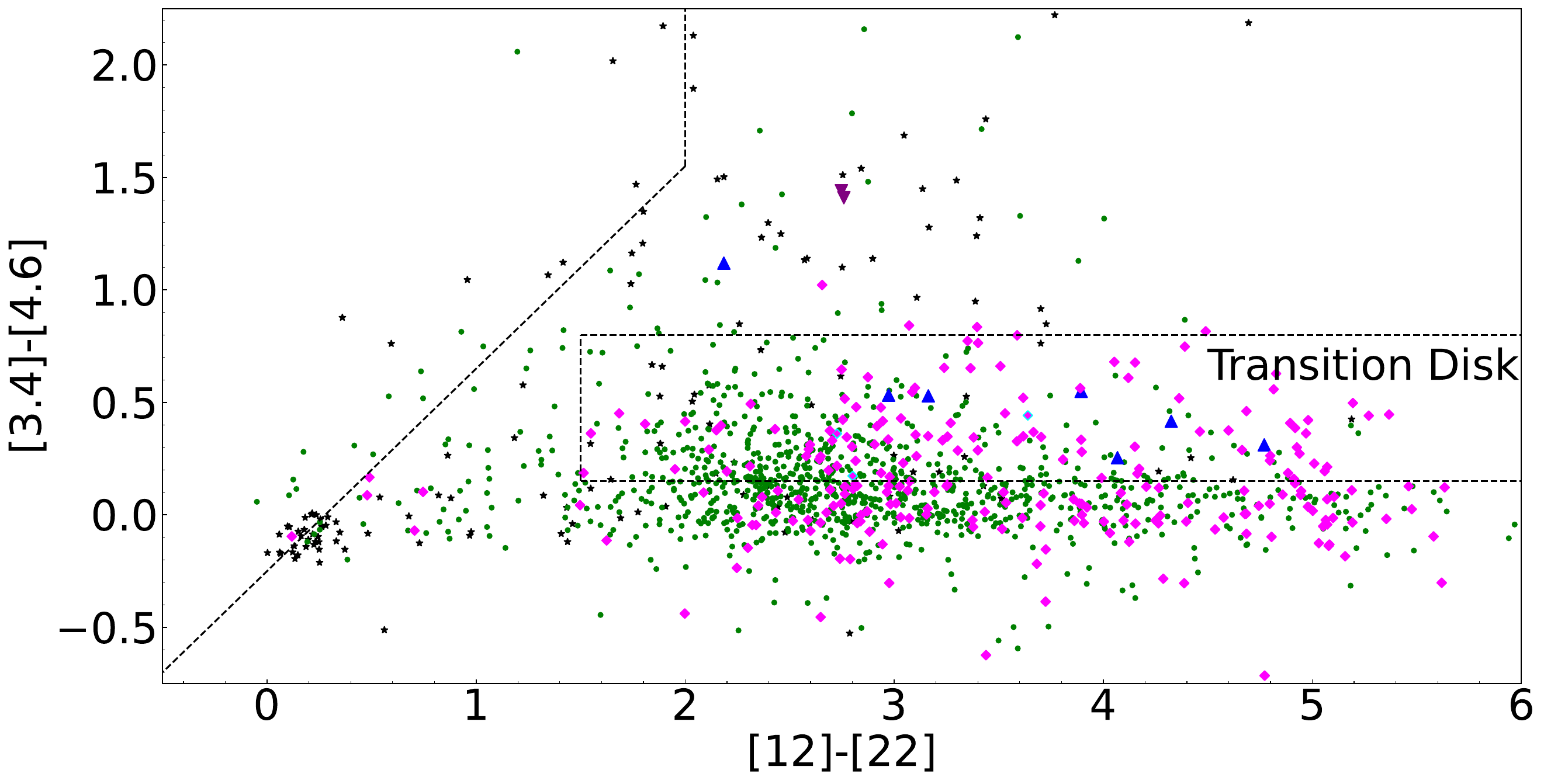} \\
\includegraphics[scale=0.18]{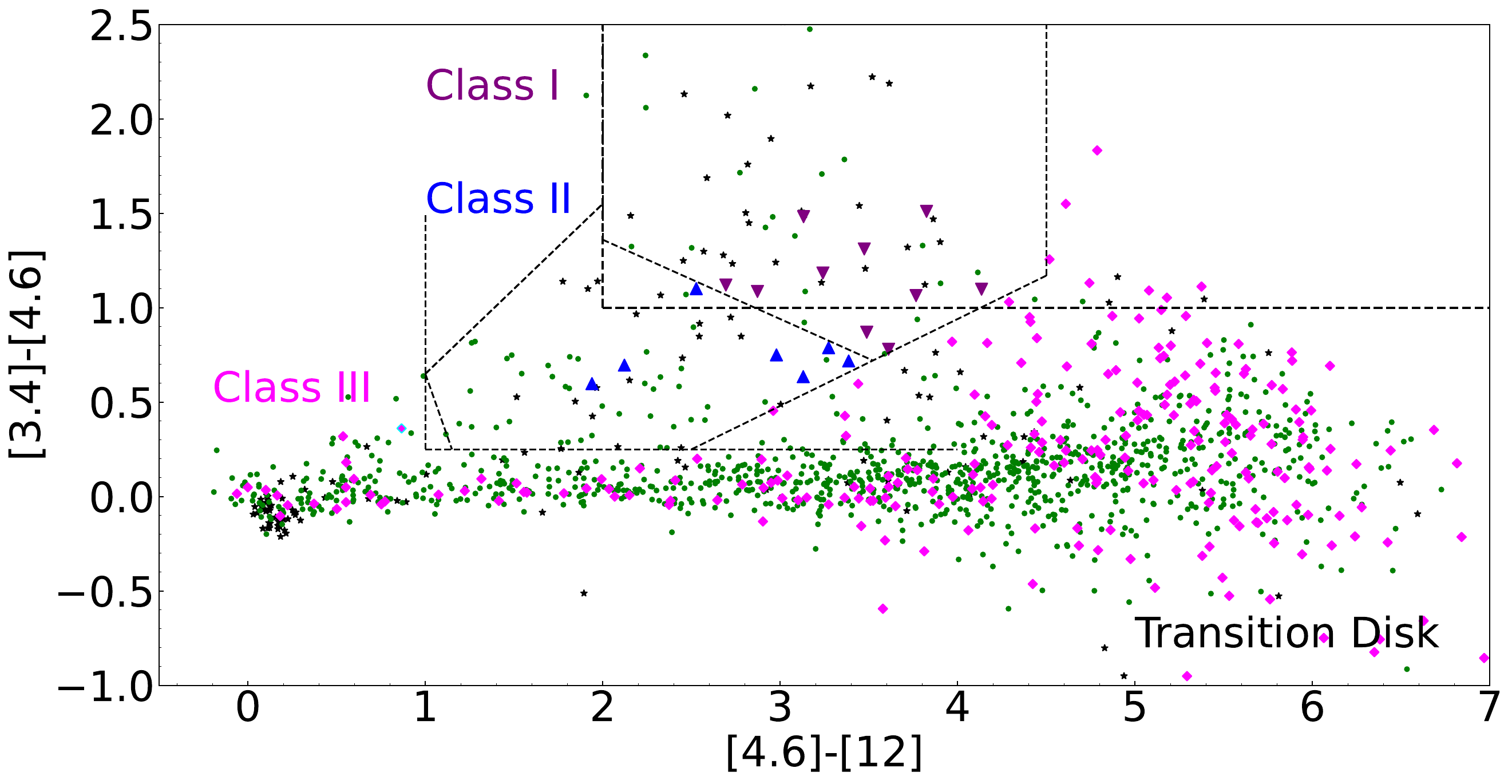}
\includegraphics[scale=0.18]{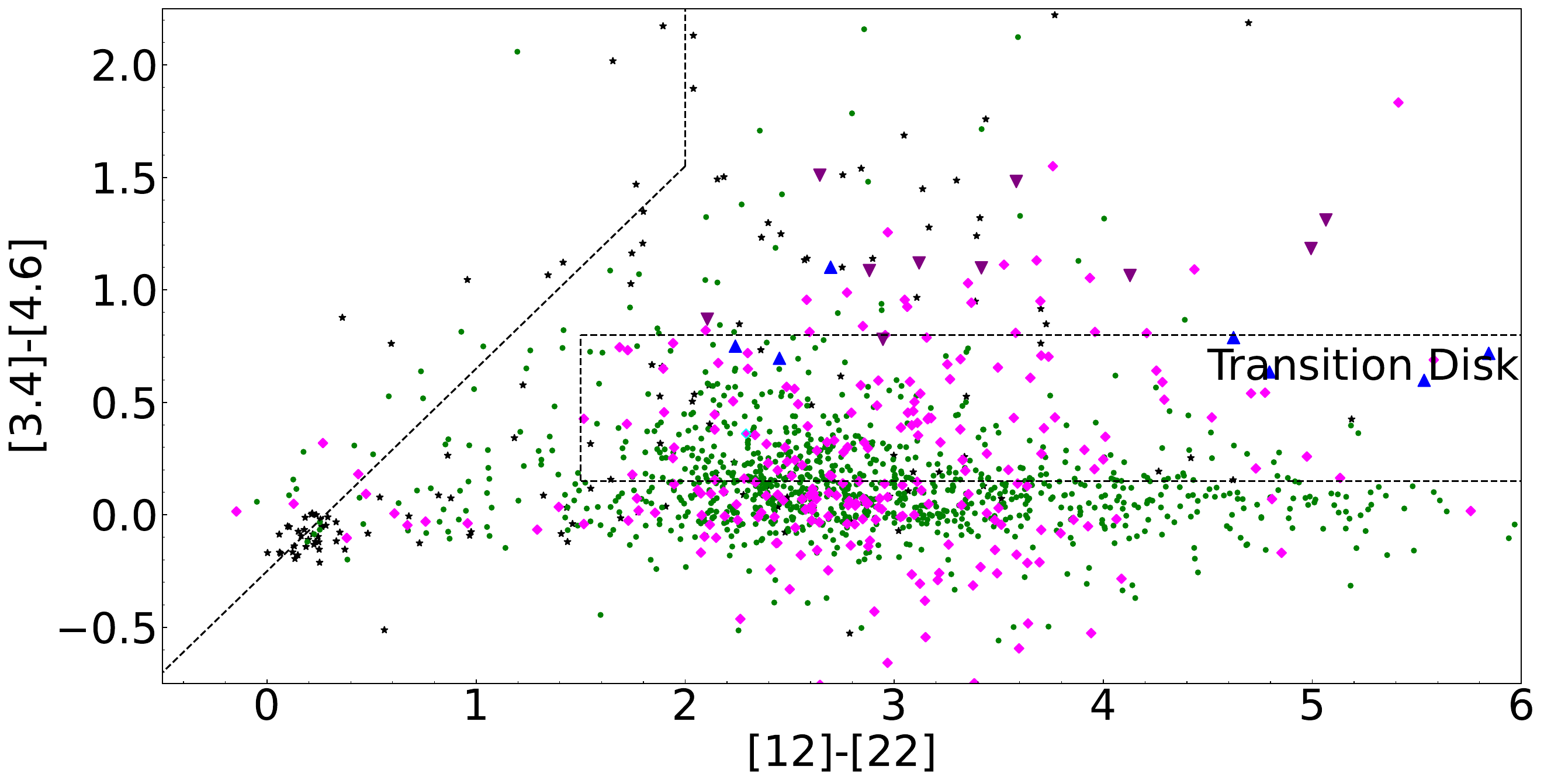}
\end{center}
\caption[width=\textheight]{Typical all--WISE color-color diagrams for LHAASO J2108+5157 (top), Kronberger~80 (middle) and Kronberger~82 (bottom). Symbols are as follows: purple triangles, blue triangles, and magenta diamonds are Class~I, Class~II, and Class~III plus transition disks YSO, respectively. Green dots are YSO sources in Cyg-OB2, while black stars are from the control field objects, both for comparison. Dashed lines show limits for YSO classes and are marked following \citet{Kang2017} and references therein. \\
Alt text: Six diagrams of photometric color combinations of the WISE mid-infrared filters of the LHASSO J2108+5157, Kronberger~80 and Kronberger~82 regions compared to the Cygnus~OB2 region (See the text for a description of the stellar counts by class).}
\label{fig:J2108_ALLWISE}
\end{figure*}

\section{Results and Discussion}
\label{sec:res_disc}

\subsection{Photometry of LHAASO J2108+5157, Kron~80 and Kron~82}
\label{sec:J2108_phot}

In Fig.~\ref{fig:Gaia_cmd} we present {\it Gaia} {\citep{gaiadr3} CM (top) and vector plot (down) diagrams for Kron~80 and Kron~82, where G stands for the G-band (centered on $\sim$430~nm) while BP (330-680~nm) and RP (640-1000~nm) for the blue and red integrated photometry data (from low-resolution photometers), respectively.
These plots include the cluster and surrounding field areas, showing little to no difference at these wavelengths.

To perform J2108 IR photometry, we consider a region with a diameter of 0.7$^{\circ}$ (about three times the upper limit of the LHAASO KM2A PSF) and centered in $\alpha_{2000}$ = 21$^{h}$08$^{m}$52.8$^{s}$, $\delta_{2000}$ = $+$51$^{\circ}$57$^{m}$00$^{s}$ \citep[see the smallest circle on top of Fig.~2, in][]{delaFuente2023b}. This region excludes Kron~82 and IRAS~21046+5110, the nearest star--forming regions \citep[see top of Fig.~1, in][]{delaFuente2023b}, since we are interested in searching for possible YSO or embedded star clusters revealed by IR photometry at J2108 (see also Fig.~\ref{fig:Krons_IR}, this paper). 

In Fig.~\ref{fig:J2108_2MASS} (top), we show the typical 2MASS CC diagram, $J-H$ vs. $H-K$, and the CM diagram $K$ vs. $H-K$, where a comparison is shown between J2108 (red dots) and Cyg-OB2  located at $\alpha_{2000}$ = 22$^{h}$33$^{m}$12.0$^{s}$, $\delta_{2000}$ = 41$^{\circ}$19$^{m}$00$^{s}$, size of 2.0$^{\circ}$, {\bf (}green dots). The blue dashed parallel lines indicate the direction of the reddening vectors considering a standard extinction law of~\citet{Rieke1985}; on them, crosses are indicating A$_{\rm V}$~=~5~mag separations. A spread in visual extinction for J2108 is seen, with very few sources above A$_{\rm V}$~=~15~mag. In addition, two stellar populations, one around A$_V$=5 and the other around A$_V$=10 magnitudes, are seen in both NIR diagrams.

From the J2108 NIR CC diagram, about 150 sources ($\sim$11\%) are young objects, Herbig Ae/Be and T-Tauri stars. These are a few sources in comparison with about 6000 ($\sim$12.4\%) for Cyg-OB2 (found to be a PeVatron source; see \S~\ref{sec:introduction}). Also, the more massive young sources, Herbig Ae/Be, are about 80 ($\sim$6.6\%) in J2108 versus about 3000 ($\sim$7\%) in Cyg-OB2, showing a lack of sources that potentially could accelerate particles via stellar winds, besides the fact that these sources do not appear clustered in J2108 (see \S~\ref{sec:cluster_analysis}).
In the J2108 NIR CM diagram, the evolutionary track corresponds to the ZAMS from \citet{Schmidt-Kaler1982} with spectral types labeled along it. A comparison between populations shows that J2108 has fewer sources in the higher mass regime, since only 12 sources ($\sim$0.8\%) have spectral type O, while Cyg-OB2 populates the region earlier than the B0 spectral type with more than 1200 sources ($\sim$2.4\%), many of them (365) above spectral type O5, implying a factor of 100 more hot massive stars in Cyg-OB2 (as a confirmed PeVatron) than in J2108, in concordance with the optical images. 

For Kron~80, NIR CC diagram reveals a lack of reddened T-Tauri stars, only 5 ($\sim$3.5\%), and more Herbig Ae/Be objects 30 ($\sim$21\%) in comparison, while for Kron~82, its CC diagram reveals the opposite, 7 ($\sim$12\%) T-Tauri stars and 5 ($\sim$8\%) Herbig Ae/Be objects. NIR CM diagrams of both clusters show a lack of massive sources, with only one O star in each one. Therefore, we conclude that none of these regions, J2108, Kron~80 and Kron~82 have enough massive stars to accelerate particles through energetic winds that could justify their possible PeVatron origin.

Nevertheless, to analyze the young stellar content in these regions, in Fig.~\ref{fig:J2108_ALLWISE} we show two typical all--WISE CC diagrams for the same areas that were used in the 2MASS diagrams. These combinations of colors are more useful than the 2MASS ones in classifying YSO, especially when no Spitzer observations (IRAC or IRAC+MIPS) are available. 

Based on these diagrams, which follow the YSO classification criteria by \citet{Koenig2014} through equations 6, 7 and 8 in \citet{Kang2017}, a summary of the YSO counts in the two clusters Kron~80 and Kron~82, and J2108 region in comparison to the PeVatron source Cyg-OB2, is shown in Table~\ref{tab:class-counts}. Note that coincidentally, the total number of YSOs in J2108, Kron~80 and Kron~82 regions are similar. However, the radius of J2108 is a factor $\sim$10 greater than the radii of Kron~80 and 82 regions since this is the limiting size of the LHAASO emission; and Kron~80 being at a distance of 5 to 6 times greater than Kron~82, J2108 and Cyg-OB2.

The canonical all-WISE diagram is the first combination of colors (in the column to the left of Fig.~\ref{fig:J2108_ALLWISE}). In contrast, transition disks are better constrained by the second combination of colors (diagrams to the right of Fig.~\ref{fig:J2108_ALLWISE}), which correspond to a stage between Class~II and Class~III objects.
In all regions, there are more transition disks and Class~III objects than Class~I and Class~II. Cyg-OB2 is more predominant as a star-forming region than J2108. 

\begin{table}[h]
    \centering
    \footnotesize
    \begin{tabular}{|l|c|c|c|c|}
    \hline
         Source & Cyg-OB2 & J2108 & Kron~80 & Kron~82 \\
         \hline
         Class~I & 19 & 14 & 2 & 10 \\
         Class~II & 40 & 7 & 7 & 7 \\
         Transition disks & 939 & 270 & 230 & 242 \\
         Class~III & 17 & 4 & 3 & 1 \\
         \hline
         Total & 1015 &  281 & 242 & 260 \\
         \hline
    \end{tabular}
    \caption{YSO classification counts from WISE data.}
    \label{tab:class-counts}
\end{table}

\subsection{Distance estimations for Kronberger~80 and 82}
\label{sec:distanceKron82}

Several distance estimates were made for Kron~80. \citet{2013A&A...558A..53K} in the Milky Way Star Cluster Catalog (MWSC), estimate a distance of 5~kpc, and \citet{Molina-Lera2019} estimated a distance of 7~kpc using distance modulus and color excess values. Most recently, \citet{2020A&A...633A..99C} estimated a distance of 10~kpc using {\it Gaia} data and machine learning methods. We adopted this last distance estimate because it also contains astrometry information, making it more reliable. There was no previous report on the distance to Kron~82 based on {\it Gaia} data.

As mentioned in \S~\ref{sec:krons}, the distance to Kron~82 is poorly constrained, with estimates ranging from 0.8 to 2.3~kpc. This study aims to determine a more accurate distance to Kron~82, following the methodology discussed by \citet{delaFuente2023c} and using data from \citet{delaFuente2023b}. To achieve this, we used the Bayesian distance calculator \citep[version 2; ][]{2019ApJ...885..131R}, which requires the celestial coordinates of the source and its systemic velocity in the Local Standard of Rest (LSR) frame. The systemic velocity of Kron~82 is reported to be V$_{\rm sys}$=$-$6.6$\pm$0.1 km s$^{-1}$ \citep{delaFuente2023b}. The resulting probability density curve for Kron~82 distance is presented in Fig.~\ref{fig:dist_calc}. As shown in the figure, the calculator provided three probability components. The component of highest probability is located at a distance of 1.62$\pm$0.05~kpc, with a probability of 0.66. The second most significant peak indicates a distance of 1.28$\pm$0.23~kpc, with a probability of 0.32. A third less significant peak is located at a distance of 0.68$\pm$0.02~kpc, with a probability of only 0.02.

\begin{figure}[!ht]
\begin{center}
\includegraphics[width=\columnwidth]{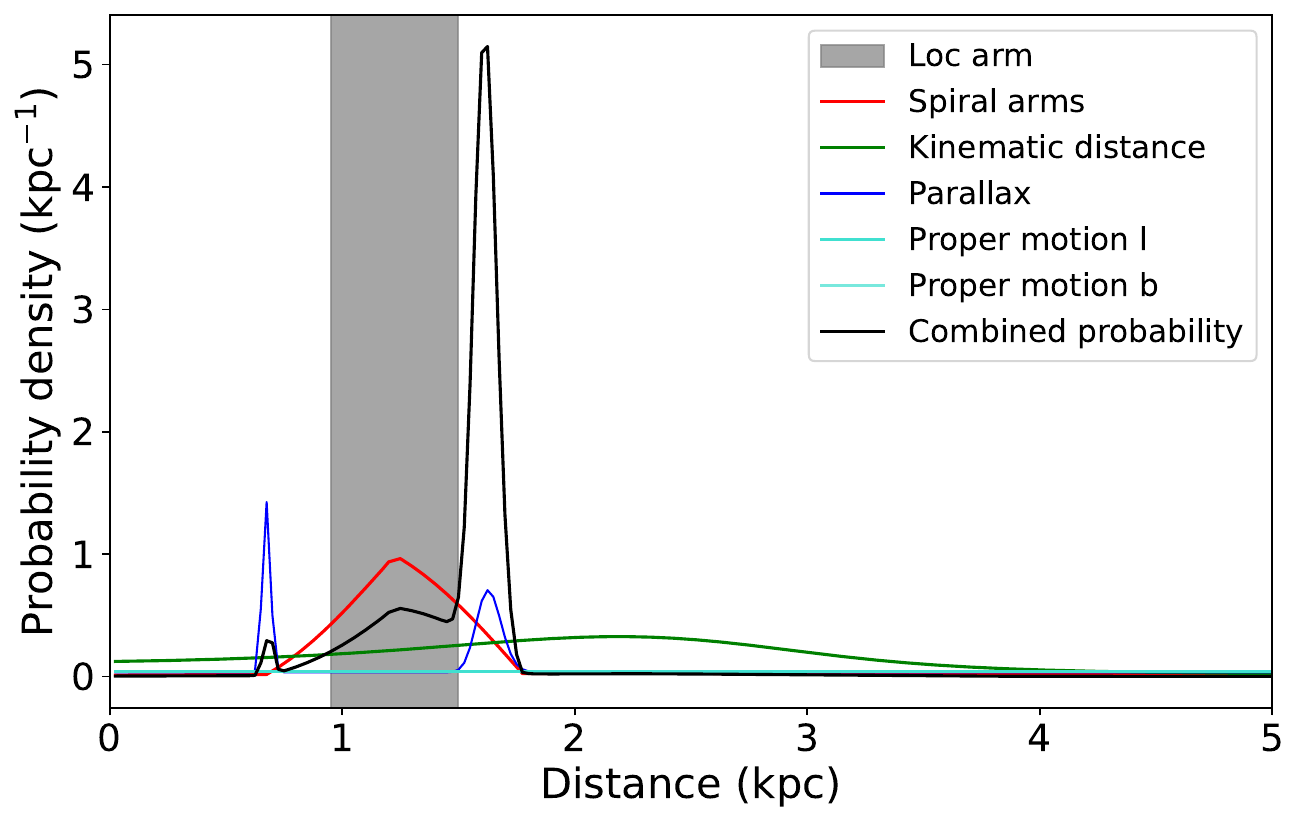}
        \end{center}
    \caption{Probability density curve for the distance of the source Kronberger~82 obtained from the Bayesian distance calculator \citep[version 2; ][]{2019ApJ...885..131R}. \\
    Alt text: Probability curve graph to identify the most probable distance to Kronberger~82.}
    \label{fig:dist_calc}
\end{figure}

Although the extinction is slightly higher due to a nebula in Kron~82, the {\it Gaia} DR3 data \citep{gaiadr3} are available to determine its distance using the following quality criteria:

\begin{equation}
 {\rm \sigma(G), \, \sigma(BP), \, \sigma(RP) }< 0.2\, {\rm mag}.
 \label{eq:Gaia}
\end{equation}

\begin{figure}
    \centering
    \includegraphics[width=\columnwidth]{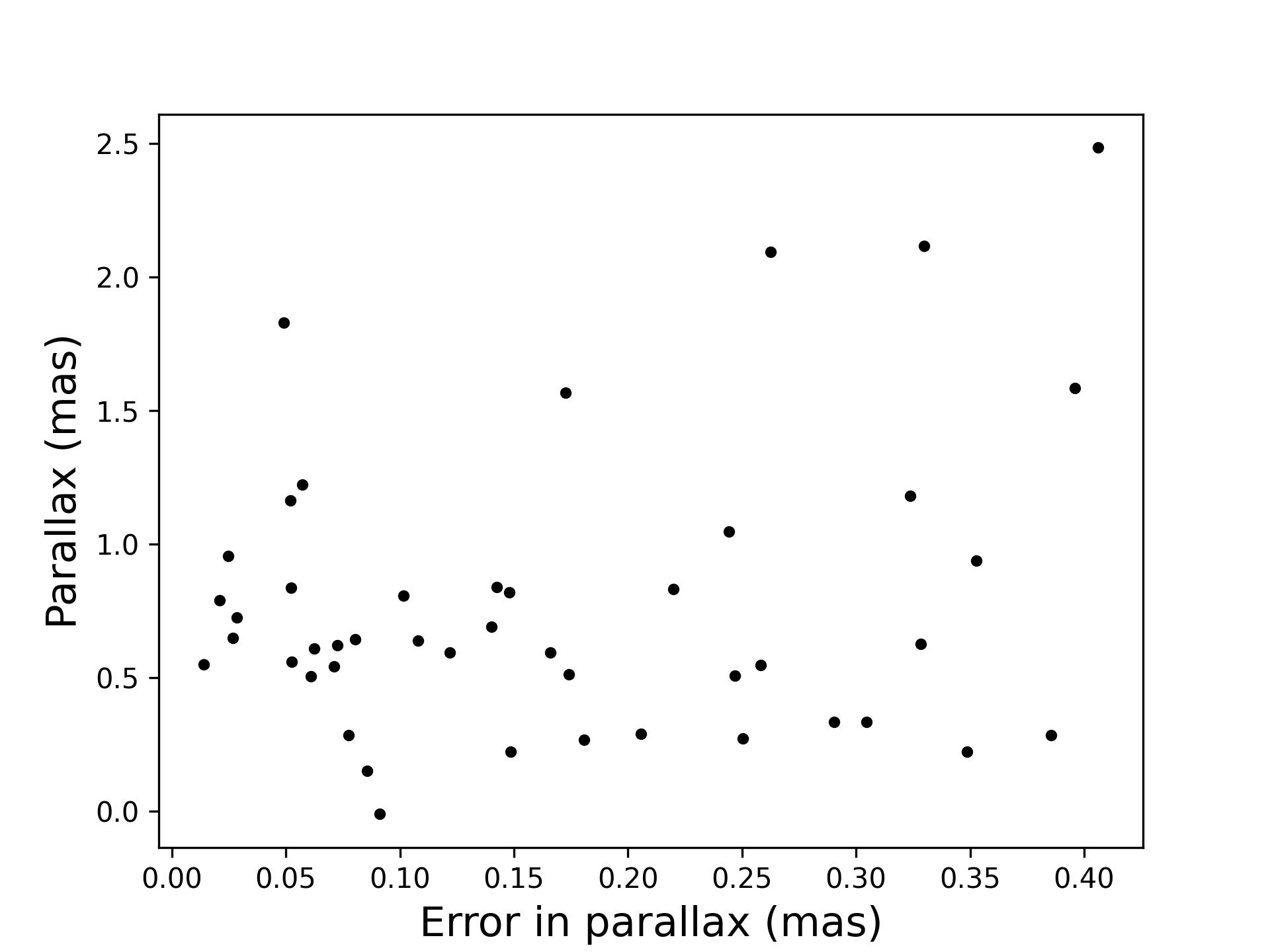}
    \caption{{\it Gaia} parallax error distribution for Kronberger~82. \\
    Alt text: Dispersion of error values in the parallax measurement of visible stars in Kronberger~82.}
    \label{fig:kron 82 parallax}
\end{figure}

The parallax error is restricted to a cut-off of 0.15~$mas$ (Fig.~\ref{fig:kron 82 parallax}) to use the most precise values in distance determination. The parallaxes are then converted to distances using the \cite{2015PASP..127..994B} code for the DR3 data. Outlier distances were removed from the stellar sample and the estimated distance was recalculated as $1.635\pm0.130$~kpc. Furthermore, the maser [HLB98] Onsala 150 \citep{2019ApJ...885..131R} is located in Kron~82, and its measured parallax from VLBI arrays is 0.613$\pm$0.020~$mas$ for a distance of 1.631$\pm$0.053~kpc. Based on these findings and our calculations, we adopt a distance of 1.63$\pm$0.05 kpc for Kron~82. 
  
Furthermore, we found an existing study of IRAS~21078+5211 HII region (within 2$\arcmin$ from Kron~82) --which is most likely associated with the Kron~82 cluster--, as an independent distance estimate of $\sim$1.6~kpc by \citet{Moscadelli2021}, which coincides with our estimate. Due to this association of the IRAS source with the cluster, we can use the reported YSO masses to estimate the whole cluster Kron~82 mass. The most massive YSO in IRAS~21078+5211 is estimated to be 6~$M_\odot$, and with the current accretion rate, the most massive star that could form in this cluster would be at least 8~$M_\odot$ \citep{Moscadelli2021}. From \citet{2004MNRAS.349..735B}, an empirical relation between the most massive star (m$_{max}$) and its total cluster mass (M$_{ecl}$) is given by:

\begin{equation}
    m_{max} = 0.39 \times M^{2/3}_{ecl}.
\end{equation}

Therefore, a cluster mass of roughly 60-100~M$_\odot$ based on the most massive YSO of 6-8M$_\odot$. 

\begin{figure}[htbp]
\begin{center}
        \includegraphics[width=\columnwidth]{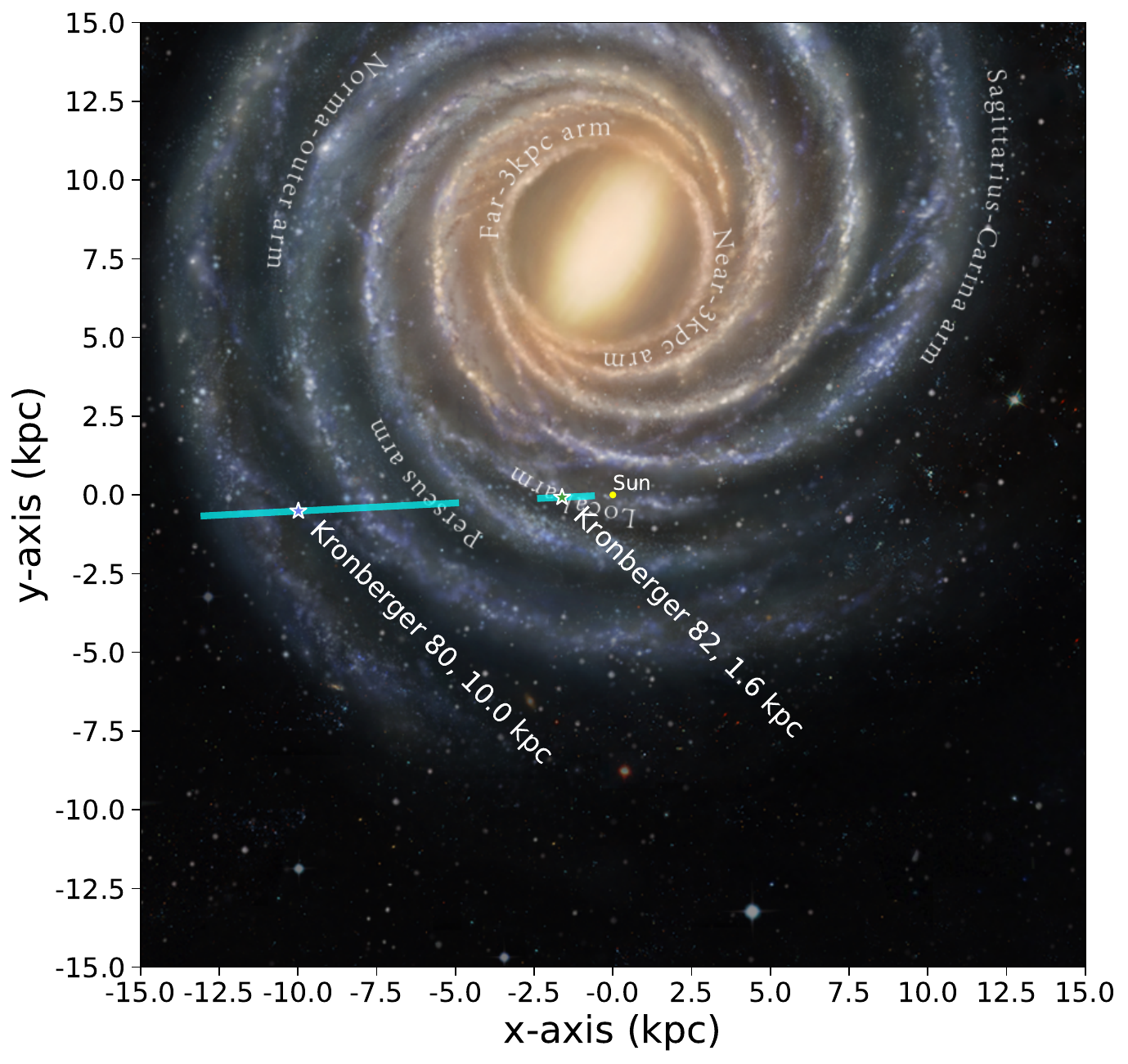}
        \end{center}
    \caption{Location of Kronberger~80 and 82 within our Galaxy. The bars indicate the estimated distance intervals for clusters. The Sun is located at position (0,0). The background image has been taken from \citet{SciAme2020}. Credit: Xing-Wu Zheng \& Mark Reid  BeSSeL/NJU/CFA. \\
    Alt text: Schematic location of Kronberger~80 and Kronberger~82 within the Milky Way. }
    \label{fig:location_Gal}
\end{figure}

The locations of Kron~80 and Kron~82 within our Galaxy are presented in Fig.~\ref{fig:location_Gal}. The bars in the figure represent the assumed distance ranges for both clusters. This illustration clearly shows that although Kron~80 (distance $\sim$10~kpc) and Kron~82 (distance $\sim$1.6~kpc) appear relatively close in projection on the sky, they are likely associated with different spiral arms, supporting the no association of Kron~80 with J2108 because of their different distances.

\subsection{Cluster Ages}
\label{sec:ages}

For Kron~80, \citet{Molina-Lera2019} determined an age between 10 and 30 Myr,  by fitting the models to the {\it ugri} and  2MASS data, but the fit appears to be poorly constrained, making the age estimate uncertain. We use deeper IPHAS data with a limiting magnitude of 18~mag. We also use our estimates of the radius and center of the cluster of 2.5$\arcmin$ and $l=92.9315^\circ$, $b=2.7963^\circ$ (\S ~\ref{sec:cluster_analysis}) to retrieve IPHAS data for Kron~80 and statistical decontamination in the Color-Magnitude space was used to remove the foreground sources. 
At a distance of 10~kpc, the PARSEC isochrones of \citet{2017ApJ...835...77M} were used to calculate the reddening of $E_{B-V}$ = 1.56. From the isochrones in Fig.~\ref{fig:kron 80 isochrones}, this value suggests a likely age between 5~Myr and 12.6~Myr (log [age] between 6.7 and 7.1), but as the errors increase toward the fainter end, we consider a mean age of 8~Myr as a most likely age of Kron~80. 

\begin{figure}[htbp]
    \centering
    \includegraphics[width=0.8\columnwidth]{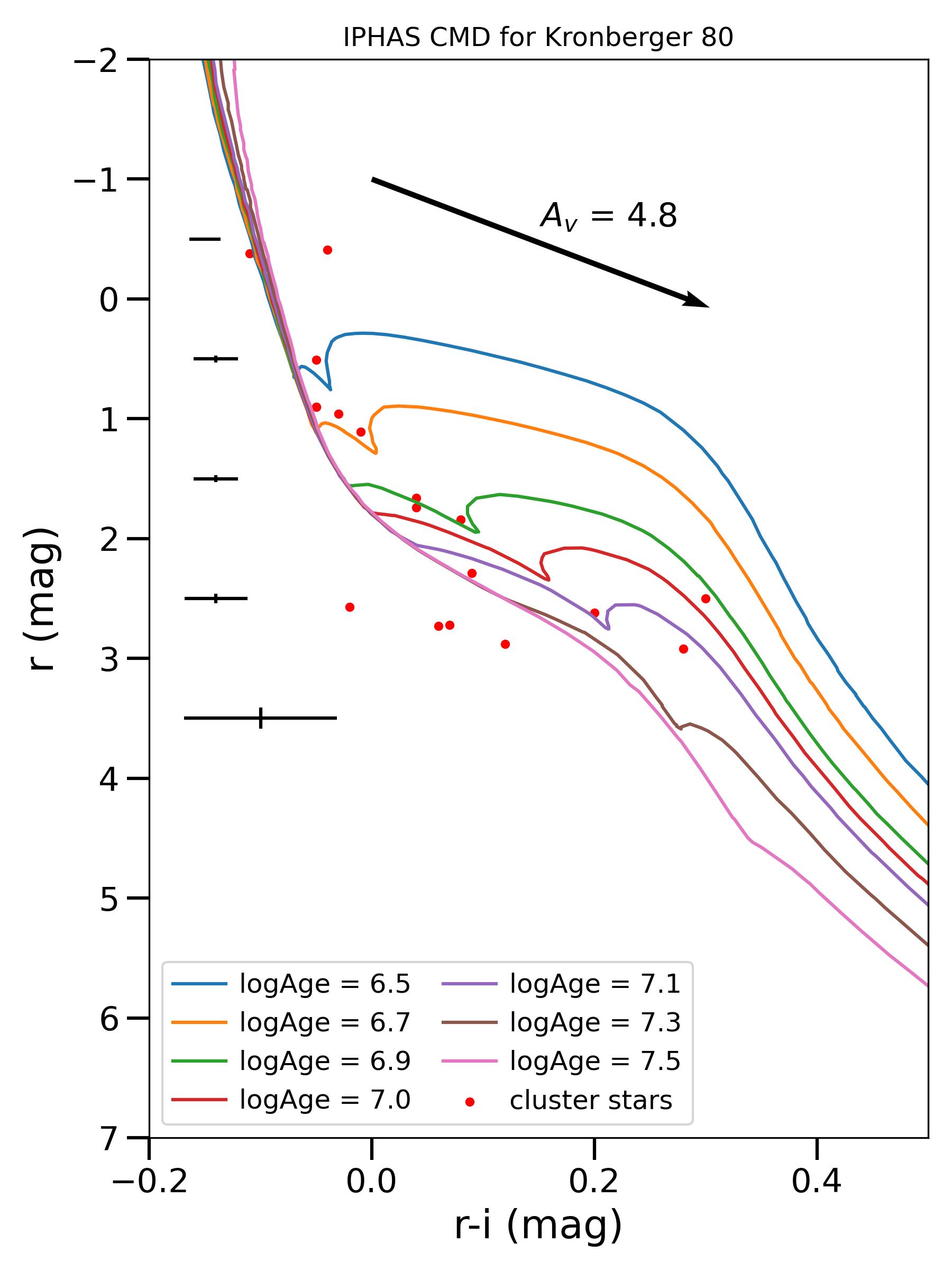}
    \caption{Optical color-magnitude diagram of Kronberger~80 derived from the INT Photometric H$\alpha$ Survey (IPHAS). The isochrones between the age range 10--30 Myr and of solar metalicity are taken from \citet{2017ApJ...835...77M} that have been corrected for cluster reddening and distance modulus. The reddening correction accounts for interstellar dust effects, ensuring that the observed colors align with the true stellar characteristics. Sample of seven isochrones and cluster stars are shown. The y-axis shows the absolute magnitude of r-band. \\ 
    Alt text: Color-magnitude diagram of optical filters of the stars in Kronberger~80. Isochrone curves at different ages are shown to visualize the estimate of the cluster age.} 
    \label{fig:kron 80 isochrones}
\end{figure}

Due to the higher extinction towards Kron~82 cluster, an age estimate using IPHAS data was not possible, and the estimation reported by \citet{Moscadelli2021} was assumed. We summarize the physical parameters of both Kronberger clusters in Table~\ref{tab:cluster_properties}.

\begin{table}[h]
    \centering
    \begin{tabular}{|c|c|c|c|c|}
    \hline
         Cluster & $l$ (deg) & $b$ (deg) & radius & Age \\
         \hline
         Kron 80 & 92.9315$^\circ$ & 2.7963$^\circ$ & 2.5\arcmin & 5-12.6 Myr \\
         Kron 82 & 92.6879$^\circ$ & 3.0468$^\circ$ & 2.0\arcmin & $<5$ Myr \citet{Moscadelli2021}   \\
         \hline
    \end{tabular}
    \caption{Parameters of Kronberger~80 and 82.} 
    \label{tab:cluster_properties}
\end{table}

\subsection{Are Kronberger~80 and Kronberger~82 PeVatrons$?$}

\citet{Mitchell2024} investigate a scenario in which a nearby ($\leq$10~kyr) young supernova remnant (SNR), located within 40–60~pc of the cloud, accelerates the cosmic rays that illuminate the molecular cloud. Although no SNR with these properties is known, an undetected and embedded SNR in the region is conceivable; perhaps an SNR from a massive star (age $\gtrsim$10$^6$ yr) in the J2108 region is producing cosmic rays, while several Class II YSOs with ages of some million years continue their evolution to a main--sequence star in hundreds of million years. Whether or not a supernova or SNR has occurred at this location, in Appendix~\ref{appendix:ap1} we present the determination of the nucleon density of the molecular gas, as the most important parameter to determine whether the LHAASO energy is reached in the molecular cloud associated to J2108.

Based on Table~\ref{tab:cluster_properties} and IR photometry, Kron~80 appears to be a distant cluster (10~kpc) with a large physical radius ($\sim$7~pc), and an age of $\sim$8~Myr, containing $\sim$240 YSO, $\sim$30 of which are Herbig~Ae/Be, with only one observed O star. It is not related to J2108 because of different distances and is not considered a PeVatron source due to the small number of massive stars producing considerable winds.
On the other hand, Kron~82 is a nearer cluster (1.6~kpc) with a smaller radius ($\sim$1~pc), an age $<$5~Myr, containing $\sim$260 YSO, only 5 of which are Herbig~Ae/Be, and also with only one observed O star. Most of the stellar population has low mass, with a rough estimate of 60-100~M$_\odot$ for the cluster mass. It is likely to be associated with J2108 because of position and distance values but is not considered a PeVatron source due to its small number of massive stars.

Thus, if we compare our results regarding the stellar content of Kron~80 and Kron~82 with Cyg-OB2 (\S~\ref{sec:J2108_phot}), and despite the recalculated distance of Kron~82, which places it in the molecular cloud Cyg-OB7  (\S~\ref{sec:J2108} and \S~\ref{sec:distanceKron82}), 
neither Kron~80 nor Kron~82 regions appear to be PeVatron candidates for J2108. 

The fact that both Kronberger regions were rejected as PeVatrons candidates for LHAASO J2108+5156, has profound implications for understanding the observed gamma-ray emission. All candidates originally proposed by \citet{Cao2021b} and discussed in the literature have now been quantitatively discarded. Therefore, a couple of ideas remain to understand the observed emission: 1. continue the search for a PeVatron counterpart in J2108, including more detailed multiwavelength studies, in the case of the presence of an old PeVatron counterpart, and 2. assume that there is no PeVatron in the region and speculate that the observed gamma rays originate from the interaction of cosmic rays produced by an external PeVatron interacting with Cyg-OB7 at the position of J2108. The first scenario may involve a supernova explosion and related objects, and the second, although very unlikely, is the identification of the nearest PeVatron. We discuss this idea in the Appendix~\ref{appendix:ap1}: for Cyg--OB2 as the closest PeVatron to J2108, a nucleon density n(H$_2$+HI) $\gtrsim$ 1.9 $\times$ 10$^{4}$ cm$^{-3}$ is needed to produce the (sub-)PeV emission.

\section{Conclusions}
\label{sec:conclusion}

We present a near- and mid-infrared analysis of the star-forming regions Kronberger~80 and Kronberger~82 to study the stellar content on them and, by comparison with the Cygnus~OB2, confirm if they can be considered PeVatrons in the vicinity of LHAASO~J2107+5158. 
Our conclusions are as follows.  

\begin{itemize}

\item Kronberger~80 is a star-forming region lying between the Perseus arm and the New arm with a distance of about 10~kpc and a radius of 2.5$\arcmin$, an age range between 5 and 12.6~Myr, and stellar content with a lack of massive stars (only one O spectral type star).     \\

\item New distance estimation for Kronberger~82, places it in the Cygnus~OB7 molecular cloud (Local Arm), at about 1.6~kpc with a radius of 2.0$\arcmin$, an age of less than 5~Myr, and stellar content mostly of low and intermediate mass (only one O spectral type star).     \\

\item In contrast to Cygnus~OB2, Kronberger~80 and 82 seem to be star-forming regions ruled by formation instead of evolution.  
Their stellar content implies the lack of stellar winds and intense UV radiation, disqualifying them as PeVatron candidates associated with the LHAASO J2108+5157 region.\\

\item LHAASO~J2108+5157 is a region in Cygnus OB7, without a noticeable clustering of stellar sources, and even its stellar content of massive stars is larger (12 O-type) than in the cluster Kronberger regions, represents only a hundredth of the same in the PeVatron Cygnus OB2. \\

\item The presence of a PeVatron in the location of LHAASO~J2108+5157 is still a mystery but it is mandatory to explain the observed (sub-)PeV emission. Considering Cygnus~OB2 as the closer PeVatron, the density of nucleons n(H$_2$+HI) in Cygnus~OB7 must be $\sim$10$^4$~cm$^{-3}$.  \\

\end{itemize}

\noindent \textbf{\large Conflict of Interest}

\noindent The authors declare that they have no conflict of interest directly relevant to the content of this article.

\begin{ack}

The authors thank the anonymous reviewer for the valuable contributions and constructive criticism that helped to improve the manuscript. This research was supported by the Inter--University Research Programme of the Institute for Cosmic Ray Research (ICRR), University of Tokyo, Grant 2024i--F--05. E de la F thanks PROSNI 2024 and the Onsala Space Observatory for financial support during a research visit in 2024, and the ICRR for research visits in March 2024, December 2024, and February 2025. He also thanks Marco P\'erez and offices of Centro Universitario de Ciencias Exactas e Ingenierias (CUCEI), Universidad de Guadalajara, for the financial support in the ICRR February 2025 research stay. 
RKY gratefully acknowledges the support from the Fundamental Fund of Thailand Science Research and Innovation (TSRI) confirmation No. FFB680072/0269, through the National Astronomical Research Institute of Thailand (Public Organization). HEV and IT--J gratefully acknowledge the financial support of the Consejo Nacional de Ciencia Humanidades y Tecnolog\'ia (CONACHyT), M\'exico; Grants 708843 and 754851. This publication presents data from the 1st Two Micron All Sky Survey, a joint project of the University of Massachusetts and the Infrared Processing and Analysis Center/California Institute of Technology, funded by the National Aeronautics and Space Administration and the National Science Foundation, 2. Wide-field Infrared Survey Explorer, a joint project of the University of California, Los Angeles, and the Jet Propulsion Laboratory/California Institute of Technology, funded by the National Aeronautics and Space Administration, and 3. Spitzer Space Telescope, operated by the Jet Propulsion Laboratory, California Institute of Technology, under a contract with NASA. 
We are grateful for the computational resources and technical support provided by the Data Analysis and Supercomputing Center (CADS) with the Leo-Atrox supercomputer at the Universidad de Guadalajara.

\end{ack}



{}

\begin{table*}[]
    \centering
    \begin{tabular}{|c|c|c|c|c|c|c|} \hline
      LHAASO &  Possible Associated & \textit{l} & \textit{b} & \textit{d} & $d_{\rm j2108}$ & References \\ 
       Source & PeVatron candidate & [deg] & [deg] & [kpc] & [pc] & \\ \hline
      J2229+5927u & -- & -- & -- & -- & -- & -- \\
      J2238+5900 & -- & -- & -- & -- & -- & -- \\
      J2228+6100u & SNR G106.3+02.7 & 106.35 & 2.71 & 0.8 & 876 & \citet{Abe2023b} \\
      J2200+5643u & -- & -- & -- & -- & -- & -- \\
      J2020+4034 & SNR G78.2+2.1 & 78.2 & 2.1 & 1.5 -- 2.6 & 409 -- 1096 & \citet{Acciari2023} \\
      J2031+4052u & Cygnus Cocoon, Cyg--OB2 & 80.22 & 0.80 & 1.4 & 388 & \citet{Cardillo2023} \\ 
      J2047+4434 & -- & -- & -- & -- & -- & -- \\
      J2031+4127u & PSR J2032+4127  & 80.22 & 1.04 & 1.8 & 410 & \citet{Bykov2024}
      \\
      \hline
    \end{tabular}

    \vspace{0.2cm}
    \caption{Sub-PeV and PeV gamma-ray sources reported in the first LHAASO catalog \citep{Cao2024} near the position of LHAASO J2108+5157. Possible associated PeVatron candidates are presented together with the galactic longitude $l$, the galactic latitude $b$ and the reported distance $d$. The (physical) distance $d_{\rm j2108}$ between the associated PeVatron candidate and LHAASO J2108+5157 ({\it l}~=~92.30$^{\circ}$, {\it b}~=~2.84$^{\circ}$) was calculated based on the angular separation between the two sources using their Galactic coordinates (see text for details).}
    \label{tab:gamma_sources}
\end{table*}


\appendix



\section*{Appendix A: LHAASO J2108+5157: Gamma-ray emission and molecular gas}
\label{appendix:ap1}

In this appendix, we calculate the minimum gas density n(H$_2$+HI) at the J2108 position required to reproduce the observed (sub-)PeV gamma-ray emission, considering the hadronic nature of cosmic rays (CRs) for the nearest PeVatron, which may also diffuse into the Cyg--OB7 region.

If no PeVatron is located near J2108, is necessary to find which PeVatron is closest to it. There are five SNRs near Cyg-OB7: HB 21, DA 530, G096.0+02.0, DA 551, and 3C434.1 \citep[see Fig. 7 of][]{Dobashi2018}. However, none of these has yet been reported as PeVatron. In Table~\ref{tab:gamma_sources}, we show nearby sub-PeV gamma-ray sources within 10 degrees of galactic longitude of the position of J2108 from \citet{Cao2024}. Here we present a reported possible PeVatron candidate with its galactic coordinates $(l,b)$, the reported distance ($d$) and the projected distance from the associated PeVatron to the position of J2108 ($d_{\rm j2108}$). The latter was calculated from the angular distance between two sources using their galactic coordinates [$(l_1,b_1),(l_2,b_2)$] as follows: $\alpha$ = cos$^{-1}$ [$\sin b_1 \sin b_2 + \cos b_1 \cos b_2 \cos(l_1 - l_2)$] and the cosine law. Concerning the position in the sky, the distance to the Sun, and the angular distance to J2108, we find that the star cluster Cyg-OB2 or the Cygnus cocoon in the Cygnus-X molecular cloud is the closest known PeVatron to J2108. On this basis, we hypothesize that the PeV CRs produced by Cyg-OB2 diffuse to the Cyg-OB7 region and cause the observed gamma-ray emission from J2108. In this case, we determine the hydrogen density of the molecular gas in J2108 as required by this hypothesis.

The total energy of the CR protons $W_{\rm p}$ and the luminosity of the gamma rays $L_{\gamma}$ produced by the interactions between the CRs and the interstellar medium (ISM) are related as follows (see Eq. (1) of \citealt{Aharonian2019})

\begin{equation}
\label{eq:A1}
W_{\rm p}(\geq 10 E)= L_{\gamma}(\geq E)~t_{\pi^0}~\eta^{-1}, 
\end{equation}

\noindent where the radiative cooling time t$_{\pi^0}$ of the CR protons due to the production of neutral pions is defined as

\begin{equation}
\label{eq:A2}
t_{\pi^0}=1.5 \times 10^{15} \bigg(\frac{n}{\rm 1\, cm^{-3}}\bigg)^{-1} \, {\rm s},
\end{equation}

\noindent and the gas density \textit{n} represents the density of hydrogen nucleons, H$_2$ and HI, in the molecular gas. The variable $\eta$ is an enhancement factor that takes into account the chemical composition of CRs and the ISM and is assumed to be $\eta = 2.6$ at $1$ PeV \citep{Peron_Aharonian_2022}. 

The energy density of CR protons above 1 PeV originating in Cyg-OB2 and present at the J2108 position is $W_{\rm p}(\geq 10 E)$, and the sub-PeV gamma-ray luminosity of J2108 observed by LHAASO is $L_{\gamma}(\geq E)$. Thus, we can compute $W_{\rm p}(\geq 10 E)$ assuming a distance to J2108 of $1.6\,{\rm kpc}$ \citep{delaFuente2023c} and the distance to Cyg-OB2 of $1.4\, {\rm kpc}$ \citep{Ackermann2011}. Since J2108 and Cyg-OB2 are separated by $12.8^{\circ}$ from center to center according to \citet{Cao2021b,LHAASO2024}\footnote{\citet{Cao2021b} report the center of J2108 at $(l,\, b) = (92.28^{\circ},\, 2.87^{\circ})$, and \citet{LHAASO2024} for Cyg-OB2 at $(l,\, b) = (79.62^{\circ},\, 1.16^{\circ})$}, the physical distance between the two objects is $388\, {\rm pc}$. According to Fig. 5 of \citet{LHAASO2024}, $W_{\rm p}(\geq 1\, {\rm PeV})_{\rm J2108}$, the energy density of CR protons above $1$ PeV of the Cyg-OB2 origin in the center of J2108, is $2.6$ times higher than the density of $3.77\times 10^{-5}\, {\rm eV}\, {\rm cm}^{-3}$ observed on Earth by the 
Icetop experiment \citep[e.g.][and references therein]{Lipari2020}. Therefore, we obtain a $W_{\rm p}(\geq 1\, {\rm PeV})_{\rm J2108}$ of about 4.0$\times 10^{-5}\, {\rm eV}\, {\rm cm}^{-3}$. 

On the other hand, the extension of J2108 is parameterized as

\begin{equation}
\label{eq:A4}
r_{\rm ext}=27.9~\bigg(\frac{\Delta \theta}{1~{\rm deg}}\bigg)\, {\rm pc},
\end{equation}

\noindent where $\Delta \theta$ is the angular extension of J2108. Therefore, assuming a spherical symmetry for the morphology of J2108, $W_{\rm p}(\geq 1\, {\rm PeV})$ is approximately calculated as 

\begin{eqnarray}
\label{eq:A5}
W_{\rm p}(\geq 1\, {\rm PeV}) &\sim& \frac{4}{3}~\pi r^3_{\rm ext} \times w_{\rm P}(\geq 1\, {\rm PeV})_{\rm J2108} \nonumber \\ &\simeq& 1.7 \times 10^{44} \bigg(\frac{\Delta \theta}{1~{\rm deg}}\bigg)^3\, {\rm erg}.
\end{eqnarray}

For the calculation of the sub-PeV gamma-ray luminosity $L_{\gamma}(E\geq 100\, {\rm TeV})$ of J2108, we refer to \citet{Cao2021b}. These authors found that the differential gamma-ray energy spectrum of J2108 follows a simple power-law function with an index of $\Gamma=-2.83$ in $20\, {\rm TeV}<E<500\, {\rm TeV}$, and calculated the gamma-ray luminosity of J2108 by:

\begin{equation}
\label{eq:A7}
L_{\rm \gamma} (20\, {\rm TeV} < E < 500\, {\rm TeV}) = 1.4 \times 10^{32}~\bigg(\frac{D}{~ 1 \rm{kpc}}\bigg)^2\, {\rm {erg ~ s^{-1}}},
\end{equation}

\noindent where $D = 1.6\,{\rm kpc}$ is the assumed distance to J2108. Considering that the gamma-ray spectrum also follows a power-law function with $\Gamma = -2.83$ above 100 TeV, this equation can be converted into $L_{\gamma}(E \geq 100\,{\rm TeV})$, the gamma-ray luminosity above 100 TeV:

\begin{equation}
\label{eq:A8}
L_{\gamma} (E \geq 100\, {\rm TeV}) = 1.0 \times 10^{32}\, \rm {erg ~ s^{-1}}.
\end{equation}

Finally, substituting Eqs.(\ref{eq:A5}) and (\ref{eq:A8}) into Eq. (\ref{eq:A1}), we obtain

\begin{equation}
\label{eq:A9}
n \sim 340 \bigg(\frac{\Delta \theta}{1~{\rm deg}}\bigg)^{-3} \, {\rm cm}^{-3}.
\end{equation}

\citet{Cao2021b} give the $95\%$ upper limit of $\Delta \theta~<$~0.26$^{\circ}$ on the extension of J2108 above 25 TeV, thus, Eq.~(\ref{eq:A9}) suggests a gas density of hydrogen $\gtrsim$~1.9~$\times$~10$^{4}$~cm$^{-3}$ at the location of J2108. 

Although this is a crude analysis and other factors such as diffuse emission or the escape time of protons have not been considered, we have found that Cyg--OB2 appears to be the closest PeVatron to J2108 and a nucleon density n(H$_2$+HI)~$\gtrsim$~1.9~$\times$~10$^{4}$~cm$^{-3}$ is required to produce the LHAASO emission by neutral pion decay. Nevertheless, \citet{delaFuente2023b,delaFuente2023c} report nucleon densities of 37 and 133 cm$^{-3}$ for FKT2022 and [FTK-MC] and \citet{Cao2021b} of n(H$_2$)~=~30~cm$^{-3}$ based on \citet{Miville2017,Dame2001}. For other regions in Cyg--OB7, \citet{Dobashi2014} found n(H$_2$) between 10$^{3}$--10$^{4}$~cm$^{-3}$ for C$^{18}$O clumps such as L1004E and IRAS~21025+5221. With average values of 10$^{2}$ to 10$^{3}$ cm$^{-3}$, it is therefore unlikely that Cyg--OB2 could be the PeVatron for J2108. The search for a PeVatron counterpart related to J2108 must continue.

\end{document}